\documentclass{aa}  

\usepackage{graphicx}
\usepackage{txfonts}
\usepackage{lipsum}
\usepackage{tikz}
\usepackage{amsmath}
\usetikzlibrary{arrows.meta, decorations.pathmorphing}
\usepackage{gensymb}
\usepackage{textcomp}
\usepackage{float}
\usepackage[colorlinks=true, allcolors=blue]{hyperref}
\usepackage{orcidlink}
\usepackage{soul}
\usepackage{lscape}
\usepackage{placeins}     

\newcommand{\citeauthorandyear}[1]{%
  \citeauthor{#1} \citeyear{#1}%
}
\begin{document}

   \title{Modeling X-ray reflection spectra from disk returning radiation: Application to 4U 1630--47}

      \author{Kostas Kourmpetis\inst{1,2}
        \and Shafqat Riaz\inst{2}
        \and Honghui Liu\inst{1}
        \and Temurbek Mirzaev\inst{3,4,5}
        \and Cosimo Bambi\inst{3,6}
        \and Debtroy Das\inst{3}
        \and Jiachen Jiang\inst{7}
        \and Kostas D. Kokkotas\inst{2}
        \and Andrea Santangelo\inst{1}}
   \institute{Institut für Astronomie und Astrophysik, Eberhard-Karls Universität Tübingen, Sand 1, 72076 Tübingen, Germany 
   \and Theoretical Astrophysics, IAAT, Eberhard-Karls Universität Tübingen, D-72076 Tübingen, Germany 
    \and Center for Astronomy and Astrophysics, Department of Physics, Fudan
University, 200438 Shanghai, China 
\and {Ulugh Beg Astronomical Institute of the Uzbekistan Academy of Sciences, Astronomy St. 33, Tashkent 100052, Uzbekistan} 
\and {Institute for Advanced Studies, New Uzbekistan University, Movarounnahr str. 1, Tashkent 100007, Uzbekistan} 
\and School of Natural Sciences and Humanities, New Uzbekistan University, Tashkent 100007,
Uzbekistan 
    \and Department of Physics, University of Warwick, Coventry CV4 7AL, UK \\[5pt]
    \email{\textcolor{black}{konstantinos.kourmpetis@uni-tuebingen.de, honghui.liu@uni-tuebingen.de}}}

\abstract
{Returning radiation is thought to play a key role in disk illumination of black-hole X-ray binaries in the high--soft state, yet it has not been fully incorporated into XSPEC reflection models. We present a new table model for reflection spectroscopy that self-consistently accounts for the returning radiation of the thermal component. To isolate this effect, we adopted the standard disk--corona configuration but disabled the corona, allowing the reflection spectrum to be produced solely by the self-irradiation of the disk. We applied our model to  three NICER and NuSTAR simultaneous observations of the black-hole X-ray binary 4U 1630--47 during the soft state. The model fits the data well, suggesting that returning radiation from the disk could potentially explain the reflection features observed in this source. Our results suggest a rapidly spinning black hole ($a_* \sim 0.99$), a disk inclination of $i \sim 43^\circ-54^\circ$, a mass accretion rate of $\dot{M}_{\rm BH} \sim (0.13-0.25)\,\dot{M}_{\rm Edd}$, and a source distance of $7.3-11.3 \,\rm{kpc}$, which is in broad agreement with previous studies.
}

   \maketitle
\nolinenumbers

\section{Introduction}
\begin{figure*}[ht]
\centering
\begin{minipage}[b]{0.48\textwidth}
\centering
\begin{tikzpicture}[scale=1, every node/.style={font=\small}, >={Latex[length=2mm]}]
\node[anchor=north west] at (-3.5,3.2) {\textbf{(a)}};
\shade[shading=radial, inner color=yellow!90!black, outer color=red!50, even odd rule]
    (0,0) ellipse (3 and 1)
    (0,0) ellipse (1 and 0.33);
    \draw[thick] (0,0) ellipse (3 and 1);       
\draw[thick] (0,0) ellipse (1 and 0.33);  
\node[below=2pt] at (0, -1.1) {Accretion Disk};
\filldraw[black] (0,0) circle (0.3);
\node[below=2pt] at (0, -0.3) {Black Hole};
\draw[thick, fill=cyan!25, opacity=0.8] (0,1.8) ellipse (0.9 and 0.5);
\node at (-1.5,2.2) {Corona};
\draw[->, thick, red] (-1.5,-0.1) -- (-2.2,1.8);
\draw[->, thick, red] (-1.9,-0.1) -- (-2.6,1.8);
\draw[->, thick, red] (-2.3,-0.1) -- (-3.0,1.8);
\draw[->, thick, red] (-1.3,-0.1) -- (-0.05,1.6);
\node[left=2pt, red] at (-2.2, 2) {Thermal};
\draw[->, thick, blue, dashed] (0,1.9) -- (0,3) node[above] {Comptonized};
\draw[->, thick, blue, dashed] (0.1,1.9) -- (1.1,3);
\draw[->, thick, blue, dashed] (-0.1,1.9) -- (-1.1,3);
\draw[->, thick, blue, dashed] (0.05,1.6) -- (1.3,-0.1);
\draw[->, thick, green!60!black] (1.55,-0.1) -- (2.2,2);
\draw[->, thick, green!60!black] (2.05,-0.1) -- (2.7,2);
\draw[->, thick, green!60!black] (2.55,-0.1) -- (3.2,2);
\node[green!60!black, above right] at (2.1,2) {Reflection};

\end{tikzpicture}

\end{minipage}
\hfill
\begin{minipage}[b]{0.48\textwidth}
\centering
\begin{tikzpicture}[scale=1, every node/.style={font=\small}, >={Latex[length=2mm]}]
\node[anchor=north west] at (-3.5,3.2) {\textbf{(b)}};
\shade[shading=radial, inner color=yellow!90!black, outer color=red!50, even odd rule]
    (0,0) ellipse (3 and 1)
    (0,0) ellipse (1 and 0.33);
\draw[thick] (0,0) ellipse (3 and 1);
\draw[thick] (0,0) ellipse (1 and 0.33);
\node[below=2pt] at (0, -1.1) {Accretion Disk};
\filldraw[black] (0,0) circle (0.3);
\node[below=2pt] at (0, -0.3) {Black Hole};
\draw[->, thick, red] (-1.5,-0.1) -- (-2.2,1.8);
\draw[->, thick, red] (-1.9,-0.1) -- (-2.6,1.8);
\node[left=2pt, red] at (-1.7, 2) {Thermal};
\draw[->, thick, violet] (-1.5, -0.1) arc[start angle=180, end angle=0, radius=2.0cm];
\draw[->, thick, violet] (-1.5, -0.1) arc[start angle=180, end angle=0, radius=1.5cm];
\draw[->, thick, brown] (1.55, -0.1) arc[start angle=-180, end angle=0, radius=-2cm];
\node[violet] at (1.1, 2.55) {Thermal};
\node[violet] at (1.1, 2.2) {Returning};
\node[brown] at (-0.8, 2.55) {Reflection};
\node[brown] at (-0.8, 2.2) {Returning};
\draw[->, thick, green!60!black] (1.55,-0.1) -- (2.2,2);
\draw[->, thick, green!60!black] (2.05,-0.1) -- (2.7,2);
\draw[->, thick, green!60!black] (2.55,-0.1) -- (3.2,2);
\draw[->, thick, green!60!black] (-2.5,-0.1) -- (-3.3,2);
\node[green!60!black, above right] at (2.1,2) {Reflection};

\end{tikzpicture}

\end{minipage}

\caption{Schematic illustration of (a) a black hole--disk corona model including thermal, Compton, and reflection components; and (b) a black-hole-disk model with a reflection component from self-irradiation of the disk (thermal returning) and iterative reflection (reflection returning). The color gradient in the disk indicates that the temperature increases toward the black hole.}
\label{fig:disk_models}
\end{figure*}

In black-hole X-ray-binary (XRB) systems, a companion star transfers matter onto a stellar-mass black hole, forming an accretion disk that emits strongly in X-rays \citep{Shakura1973, Novikov1973blho.conf..343N}. Determining the spin of stellar-mass black holes is crucial, as it provides insight into the collapse of massive stars and may help explain high-energy phenomena such as relativistic jets \citep[e.g.,][]{Penrose-1969, Blandford_Znajek-1977, King_Kolb-1999, Miller_Miller-2015}. Several techniques have been developed to measure spins, including the continuum-fitting method \citep{Zhang-1997, Mcclintock-2014}, X-ray reflection spectroscopy \citep{Fabian-1989, Reynolds-2014, Bambi-2021}, quasi-periodic oscillations \citep{Motta2014MNRAS.437.2554M, Motta2014MNRAS.439L..65M, Ingram_Motta-2019, Liu2021ApJ...909...63L}, and gravitational-wave detections \citep{Abbot-2016, Abbot-2019}. Among these, reflection spectroscopy stands out as a particularly powerful diagnostic tool, as it probes the innermost regions of accreting systems \citep[e.g.,][]{Fabian2012MNRAS.424..217F, Garcia2015ApJ...813...84G, Walton2016ApJ...826...87W, Liu2023ApJ...950....5L, Liu2023ApJ...951..145L} and enables tests of general relativity in the strong-field regime \citep{De-Rosa-2019, Reynolds-2021, Bambi-2021, Tripathi-2021a, Tripathi-2019, Liu2019PhRvD..99l3007L,Liu2020ApJ...896..160L}.

As shown in Fig.~\ref{fig:disk_models}a, the observed spectrum of black-hole XRBs typically consists of three primary components: (i) a thermal component originating from a geometrically thin, optically thick accretion disk \footnote{An accretion disk is considered geometrically thin when $h/r \ll 1$, 
where $h$ is the disk scale height at radius $r$; and optically thick if $h \gg \lambda$, 
with $\lambda$ denoting the photon's mean free path.
} \citep{Shakura1973,Page-Thorne-1974}; (ii) a Comptonized component produced by a hot corona\footnote{The corona is some hot plasma ($\sim$~100 keV) in the region surrounding the black hole, and its precise geometry remains uncertain (see e.g., \citeauthorandyear{Bambi-2017} and
\citeauthorandyear{Li-2025} for a discussion and evaluation of different geometries).} that inverse-Compton scatters soft disk photons \citep{Thorne1975ApJ...195L.101T, Shapiro1976ApJ...204..187S}; and (iii) a reflection component resulting from the illumination of the disk by the corona \citep{George1991MNRAS.249..352G}.  
The disk is thought to be cold as it can efficiently emit radiation and its thermal spectrum peaks in the soft X-rays ($\sim$~1 keV). The coronal spectrum takes a power-law-like form that can extend to above 100 keV.

In the rest frame of the accretion disk's material, the reflected X-ray spectrum typically features narrow fluorescent emission lines at lower energies and a pronounced Compton hump peaking between 20 and 40 keV \citep{Ross_Fabian-2005,Garcia-Kallman-2010}. The most notable spectral line is generally the iron $\mathrm{K\alpha}$ complex, which appears at 6.4 keV for neutral or weakly ionized iron, and shifts up to 6.97 keV for fully ionized, hydrogen-like iron\footnote{As the ionization increases, reduced electron shielding results in a higher effective nuclear charge, shifting the line energy upwards.}. However, when observed from a distant point, the disk’s entire reflection spectrum appears smeared due to relativistic effects \citep{Fabian-1989, Laor1991ApJ...376...90L, Fabian2000, Dauser2010}.

X-ray reflection spectroscopy has advanced significantly in recent years. The current state-of-the-art is represented by relativistic models such as \texttt{relxill}\footnote{\url{https://www.sternwarte.uni-erlangen.de/~dauser/research/relxill/}} \citep{Dauser-2013, Garcia-2014}, \texttt{reltrans}\footnote{\url{https://github.com/reltrans/reltrans}} \citep{Ingram-2019}, \texttt{reflkerr} \citep{Nied_Zycki-2008, Nied-2019}, and \texttt{kyn}\footnote{\url{https://projects.asu.cas.cz/stronggravity/kyn}} \citep{Dovciak-2004}, as well as radiative-transfer models such as \texttt{xillver} \citep{Garcia-2013} and \texttt{reflionx}\footnote{\url{https://github.com/honghui-liu/reflionx_tables}} \citep{Ross_Fabian-2005}, which compute the reflection spectra in the rest frame of the material of the disk. 

While these frameworks represent major improvements over earlier generations, they still rely on simplifying assumptions concerning the disk structure, coronal geometry, and the treatment of relativistic effects \citep[e.g.,][]{Bambi-2021, Liu2025MNRAS.536.2594L, Huang2025ApJ...989..168H}. Some simplifications are relatively benign. For example, the assumption of infinitesimally thin Keplerian disks often provides accurate results for rapidly rotating sources \citep{Shashank-2022,Abdikamalov-2020, Tripathi-2020, Tripathi-2021b, Jiang-2022}. However, such simplifications can introduce substantial systematic uncertainties if applied to systems near the Eddington accretion limit, since the disk becomes progressively thicker and therefore cannot be described via the thin-disk model \citep{Riaz-2020a, Riaz-2020b}.

A particularly important effect in this context is returning radiation (or self-irradiation; purple and brown arrows in Fig.~\ref{fig:disk_models}b), i.e., disk emission that is bent back onto the accretion disk by the strong gravitational field of the black hole. The first calculations of this effect were performed by \cite{Cunningham-1976} in the context of thermal disk spectra, followed by early theoretical developments by \cite{Agol-2000}; \cite{Li-2005} later showed that it can mimic the effect of a higher accretion rate. Although usually neglected in thermal spectral fitting, returning radiation has been shown to significantly influence polarization signatures \citep{Schnittman-2009}.

The role of returning radiation in reflection spectra was first examined in \cite{Dabrowski-1997}, whose authors concluded it is negligible if the corona corotates with the disk. \cite{Nied_Zycki-2008} confirmed this but showed that the effect becomes significant for a static corona close to the black hole. Subsequent studies in the lamppost geometry \citep{Nied-2016,Nied-2018} demonstrated that returning radiation can significantly alter the reflection spectrum under certain conditions. More recent work, employing a range of approximations, has further investigated this effect \citep{Wilkins-2020, Dauser-2022,Riaz-2021,Riaz-2023}. These studies generally agree that returning radiation is especially relevant for rapidly spinning black holes $(a_* \geq 0.9)$ with compact coronae illuminating the innermost disk (e.g., lamppost heights of $h < 5\ r_g$, where $r_g$ is the gravitational radius).

Different approaches have been proposed to incorporate returning radiation in reflection modeling. For instance, \cite{Dauser-2022} introduced a modified emissivity profile within \texttt{relxill}, though still assuming the illuminating flux follows a simple cutoff power law in the disk's rest frame. However, because the spectrum of returning radiation itself has the form of a reflection spectrum rather than a power law, its effect cannot always be absorbed into emissivity modifications \citep{Mirzaev-2024b}. In fact, for systems where the disk extends very close to the event horizon, the corona predominantly illuminates the innermost disk, and the ionization parameter is low to moderate; returning radiation can substantially reshape the reflection spectrum. In such cases, computing the rest-frame reflection spectrum at each disk radius using the true incident spectrum (a mix of coronal power-law and reflected spectra) becomes essential. Otherwise, spectral fits may fail, leading to significantly biased parameter estimates \citep{Mirzaev-2024b}.

On the observational side, the importance of returning radiation has only recently been recognized. The first evidence that returning disk radiation can itself generate observable reflection signatures was presented by \cite{Connors-2020} based on RXTE observations of XTE J1550–564. This result demonstrated that reflection features can arise even in the absence of a strong coronal component, opening a new avenue for interpreting soft-state spectra.
This work motivated a growing number of observational studies employing models, for example \texttt{relxillNS} \citep{Garcia-2022}, to search for similar signatures in other sources in the soft state, including 4U 1630--47 \citep{Connors-2021}, EXO 1846--031 \citep{Wang-2021}, MAXI J0637--430 \citep{Lazar-2021}, MAXI J1631--479 \citep{Rout-2023}, MAXI J1803--298 \citep{Adegoke-2024}, V4641 Sgr \citep{Connors-2025}, and GRS 1739--278 \citep{Zhao-2026}, among others. The implementation of returning radiation for the hard state was presented by \cite{Dauser-2022} through modifications to the emissivity profile, and such a model has been tested in several XRBs \citep[GX 339--4, MAXI J1535--571, and Swift J1658.2--4242;][]{Huang-2025}. While these studies typically rely on approximate treatments of self-irradiation, they collectively establish an observational foundation for the role of returning radiation, bridging earlier theoretical work and the need for more self-consistent modeling.

In this work, we built on this progress by undertaking a fully self-consistent treatment of returning radiation in the absence of a corona. Despite the growing theoretical and observational interest in this effect, there is currently no XSPEC model that self-consistently accounts for returning radiation of the thermal spectrum and its reflections. In the high--soft state, self-irradiation of the disk is expected to dominate and may reveal novel observational signatures in the reflection spectrum. The primary aim of this study is therefore to construct a table model that captures the effect of returning radiation on disk reflection in a pure-disk scenario and to test its performance by fitting real observational data of sources in the soft state. 

This paper is structured as follows. In Sect.~\ref{sec:mod} we describe the details of the physical model and the code utilized, alongside its implementation in XSPEC to generate the corresponding table model. Section~\ref{sec:analys} presents the spectral analysis of 4U 1630--47 using both our model and the widely used \texttt{relxillNS}, including the fitting procedure and interpretation of the results. Finally, Sect.~\ref{sec:concl} summarizes the main conclusions of this work and outlines possible directions for future studies. Throughout the analysis, we adopted the $u_\mu u^\mu=-1$ convention for timelike four-velocities, we used Boyer-Lindquist coordinates for the Kerr metric and employed natural units $(c = G = \hbar = k_B = 1)$, unless stated otherwise.

\section{Model development} \label{sec:mod}
\subsection{The \texttt{ziji} code}

We focused on modeling the reflection spectrum in the absence of any coronal emission. We adapted the \texttt{ziji} code (available on GitHub\footnote{\href{https://github.com/ABHModels/ziji/}{\textcolor{magenta}{https://github.com/ABHModels/ziji/}}}) presented in \cite{Mirzaev-2024a} and disabled the corona by setting its luminosity to zero, thereby isolating the effect of returning radiation on the reflection spectrum.

The default space-time metric in our model is the Kerr solution \citep{Kerr-1963}. However, \texttt{ziji} does not rely on any specific equation or property of the Kerr metric, allowing us to adopt any stationary, axisymmetric, and asymptotically flat space time by simply modifying the metric coefficients \citep{Mirzaev-2024a}. 

The accretion disk is described by the Novikov--Thorne disk \citep{Page-Thorne-1974, Novikov1973blho.conf..343N}. The disk is assumed to be infinitesimally thin and oriented perpendicular to the black hole’s spin axis, with its inner edge located at the innermost stable circular orbit $(R_{\rm isco})$, while its back-reaction to the space-time metric is negligible \citep{Page-Thorne-1974, Bambi-2024b}. Material within the disk follows nearly geodesic, equatorial, and circular trajectories. The time-averaged energy flux from the disk at radius $r$ is \citep{Page-Thorne-1974,Bambi-2017}
\begin{equation}
F(r) = \frac{\dot{M}_{\rm BH}}{4\pi M_{\rm BH}^2} \mathcal{F}(r),
\end{equation}
where $\dot{M}_{\rm BH}$ is the mass-accretion rate and $M_{\rm BH}$ is the black-hole mass. $\mathcal{F}(r)$ is a dimensionless function dependent on the specific angular momentum, $L_z$; energy, $E$; and angular velocity, $\Omega$ of the orbiting gas:
\begin{equation}
\mathcal{F}(r) = - \frac{\partial_r \Omega}{(E - \Omega L_z)^2}\frac{M_{\rm BH}^2}{\sqrt{-G}} \int_{r_{\rm in}}^r (E - \Omega L_z) (\partial_\rho L_z) \, d\rho,
\end{equation}
where \citep[see, e.g.,][]{Bambi-2017}:
\begin{align}
\Omega &= \frac{-\partial_r g_{t\phi} \pm \sqrt{ (\partial_r g_{t\phi})^2 - (\partial_r g_{tt})(\partial_r g_{\phi\phi}) }}{ \partial_r g_{\phi\phi} }, \label{eq:omega} \\
E &= -\frac{g_{tt} + \Omega g_{t\phi}}{ \sqrt{ -g_{tt} - 2\Omega g_{t\phi} - \Omega^2 g_{\phi\phi} } }, \label{eq:E} \\
L_z &= \frac{g_{t\phi} + \Omega g_{\phi\phi}}{ \sqrt{ -g_{tt} - 2\Omega g_{t\phi} - \Omega^2 g_{\phi\phi} } }. \label{eq:L}
\end{align}
Here, $G$ is the determinant of the near equatorial plane metric and
$r_{\rm in}$ is the inner-edge radius of the accretion disk
$(r_{\rm in} \geq R_{\mathrm{isco}})$. In the expression of $\Omega$ the $+$ sign is for corotating orbits and the $-$ sign is for counterrotating orbits.

The disk's effective temperature is derived via the Stefan--Boltzmann law, $F = \sigma T_{\rm eff}^4$, and corrected for nonthermal effects using a hardening (or color) factor $f_\text{col}$:
\begin{equation}
T_{\rm col} = f_{\rm col} T_{\rm eff}.
\end{equation}

The specific intensity of the radiation in the rest frame of the material in the disk is \citep[e.g., ][]{Shimura-1995, Li-2005}:

\begin{equation}
    I=\frac{2h\nu^3_{\rm e}}{c^2}\frac{1}{f_{\rm col}^4}\frac{\Upsilon}{e^\frac{h\nu_{\rm e}}{k_{\rm B} T_{\rm col}}-1},
\end{equation}
where $\nu_{\rm e}$ is the frequency of the emitted photons, $h$ is Planck’s constant, $c$ is the speed of light, $k_{\rm B}$ is the Boltzmann constant, and 
$\Upsilon = \Upsilon(\vartheta_{\rm e})$ is a function that accounts for the angular dependence of the emission. In the Kerr space time, the thermal emission at each radius is determined by the black-hole mass, $M_{\rm BH}$; mass-accretion rate, $\dot{M}_{\rm BH}$; spin, $a_*$; color-correction factor, $f_{\rm col}$ (fixed at 1.7 throughout the analysis); and angular factor, $\Upsilon$ (set to 1 for isotropic emission), as also noted in \cite{Mirzaev-2024a}. The \texttt{ziji} code also treats iron abundance as a free parameter. Throughout all subsequent modeling, the iron abundance was fixed to the solar value $(A_{\rm Fe}=1)$.

The ionization, $\xi,$ at each radius, $r,$ is given by
\begin{equation}
\xi(r) = \frac{4\pi \Phi_{\rm X}(r)}{n_{\rm e}(r)}, \label{eq:xsi}
\end{equation}
where $\Phi_{\rm X}(r)$ is the incident flux from returning radiation and $n_{\rm e}(r)$ is the electron number density, which was assumed to be constant in this work.

The luminosity of the accretion disk is
\begin{equation}
    L_{\rm disk}=\eta{\dot{M}_{\rm BH}c^2},
    \label{eq:ldisk}
\end{equation}
where $\eta=1-E_{r_{\rm in}}$ is the efficiency of the system and $E_{r_{\rm in}}$ is the conserved specific energy
calculated at the radius of the inner edge of the disk, $r_{\rm in}$ \citep{Bambi-2017}. We note that we modified the \texttt{ziji} code to accept the mass accretion rate as an input parameter instead of the luminosity (using Eq.~\ref{eq:ldisk}), as our primary interest lies in exploring variations in $\dot{M}_{\rm BH}$.

To compute the reflection spectrum in the rest frame of the disk, we used the \texttt{reflionx} code\footnote{We employed the \texttt{reflionx} code rather than the \texttt{reflionx} table, since the latter one assumes that the disk is illuminated by a power-law spectrum with a high-energy cutoff \citep{Mirzaev-2024b}.} \citep{Ross_Fabian-2005}. The radiation field incident on any point on the disk in our model includes only the following:
returning thermal radiation (purple arrows in Fig.~\ref{fig:disk_models}b;
zeroth iteration); and higher order returning reflection radiation up to three iterations\footnote{To
avoid confusion, we note that in the papers where the \texttt{ziji} code
was presented \citep{Mirzaev-2024a, Mirzaev-2024b}, up to four iterations were
performed. The zeroth iteration corresponds to the reflection component produced
by the corona. In our case, there is no corona; therefore, we performed one
fewer iteration (0–3). Physically, the procedure is identical.} (brown
arrow in Fig.~\ref{fig:disk_models}b; first iteration).

Finally, once the spectra in the rest frame of the disk in every radial bin were calculated, the relativistic spectra as seen by a distant observer were computed via ray-tracing. The observed spectral flux is given by \citep[e.g., ][]{Dovciak-2004}
\begin{align}
    F_{\rm o}(E_{\rm o}) &= \int_{source} I_{\rm o}(E_{\rm o}, X, Y) \, d\Omega \\
             &= \frac{1}{D^2} \int_{source} g^3 \, I_{\rm e}(E_{\rm e}, r_{\rm e}, \theta_e) \, 
             dX \, dY, \label{eq:ray_trace}
\end{align}
where the subscripts $o$ and $e$ stand for observed and emitted, respectively. $D$ is the distance between the observer and the source, $g=E_{\rm o}/E_{\rm e}$ is the redshift factor, and $I_{\rm o}=g^3I_{\rm e}$, following Liouville's theorem \citep[see][for more details]{Lindquist-1966}. $X$ and $Y$ are the Cartesian coordinates in the plane of the distant observer, and $d\Omega=dXdY/D^2$ is the respective infinitesimal solid angle. Lastly, $r_{\rm e}$ denotes the radius at which the photon is emitted in the accretion disk, while $\theta_{\rm e}$ represents the emission angle measured in the rest frame of the disk material (see the discussion in Sect.~\ref{sec:zj_vs_NS}). 

\subsection{XSPEC table model(s) generation}

To construct the XSPEC table model, we computed the blackbody and reflection spectra over a grid of parameters, as listed in Table~\ref{tab:xspec_parameters}. Our table model includes the following five parameters:

\begin{enumerate}
    \item electron density $(n_{\rm e})$
    \item dimensionless spin parameter $(a_*)$
    \item black-hole mass $(M_{\rm BH})$    \item mass-accretion rate $(\dot{M}_{\rm BH})$    \item inclination angle $(i)$
\end{enumerate}

The physical distance of the source is absorbed inside the normalization of the model, which is automatically introduced by XSPEC (see Appendix \ref{sec:dist}). As an initial step, we calculated the spectra in the rest frame of the disk, since including the inclination angle in the full grid would substantially increase the computational time because the code performs ray-tracing in all iterations and we are only interested in the last one. Once the rest-frame spectra were calculated, we performed ray-tracing for the third iteration only, in order to obtain the spectra in the observer’s rest frame.

\begin{figure*}[ht]
    \centering
    \includegraphics[width=0.9\linewidth]{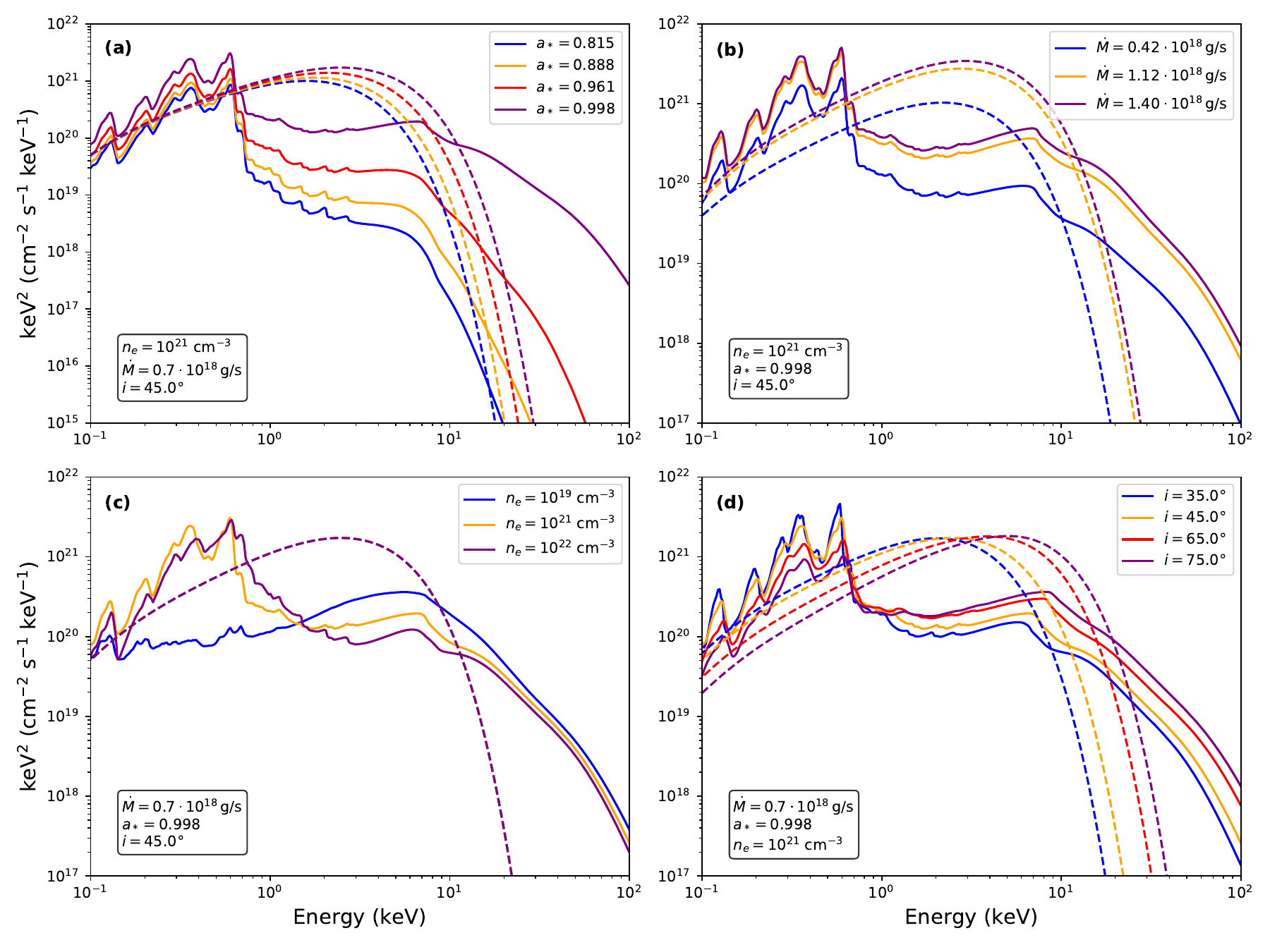}
    \caption{Reflection spectra (solid lines) produced by returning radiation (third iteration) and blackbody spectra (dashed lines) from the \texttt{ziji} code for a $10{\rm \, M_\odot}$ black hole, showing the impact of spin (a), mass-accretion rate (b), electron density (c), and inclination angle (d) on the spectra.}
    \label{fig:ziji_model_plot}
\end{figure*}

The full grid contains 9,100 points and was chosen based on parameter values commonly reported in the literature for the X-ray binary 4U 1630--47 during the high-soft state, which serves as the test case for our model (see Sect. \ref{sec:analys}). The electron density $(n_{\rm e})$ spans from $10^{18}$ to $10^{22} \, \mathrm{cm^{-3}}$; lower densities would yield high ionization parameters (see Eq.~\ref{eq:xsi}), significantly weakening the reflection features. This is illustrated in Fig.~\ref{fig:ziji_model_plot}c where even for $n_{\rm e}=10^{19} \, \mathrm{cm^{-3}}$ the reflection features are nearly absent\footnote{At high ionization parameters the disk's surface becomes highly ionized, reducing photoelectric opacity and suppressing fluorescent line production. As a result, characteristic reflection features, particularly the Fe K$\alpha$ line and absorption edges, are weakened, and the spectrum becomes dominated by a smooth Compton-scattered continuum.}.
The mass accretion rate $(\dot{M}_{\rm BH})$ covers a range of $(0.14-2.1) \, 10^{18} \, {\rm g/s}$. For reference, for a maximally spinning $10 \, {\rm M_\odot}$ black hole, a mass accretion rate of $1.4 \cdot 10^{18} \, {\rm g/s}$ corresponds to approximately $30\%$ of the Eddington accretion rate $(\dot{M}_{\rm Edd}=L_{\rm Edd}/\eta c^2)$. This value is typically taken as the upper limit of validity of the thin-disk approximation, as higher accretion rates lead to geometrically thick disks that are no longer well described by the Novikov--Thorne model \citep{Riaz-2020a,Riaz-2020b}. The black-hole spin parameter $(a_*)$ ranges from approximately $0.8$ up to the Thorne limit $(0.9982)$, while the mass $(M_{\rm BH})$ is fixed to $10 \,  {\rm M_{\odot}}$. Lastly, the inclination angle $(i)$ was permitted to vary between $30^\circ$ and $75^\circ$. We omitted a corona from this setup for two reasons. First, including lamppost corona would add two extra dimensions (luminosity and height) to the current grid. This mult-dimensional expansion would exponentially increase the grid size, necessitating a substantial amount of extra work and leading to significantly higher computational costs and processing times to calculate the new model grid. Second, the entire model was compiled into a single FITS file for use as an XSPEC table model; adding more parameters would inflate the file size beyond the practical limits of standard XSPEC table formats. Consequently, extending the model to include corona is deferred to a dedicated follow-up project.  

The effect of each of the model's parameters on the spectra is illustrated in Fig.~\ref{fig:ziji_model_plot}, where the blackbody and reflection (3rd iteration) spectra are plotted for different values of the spin (a), mass-accretion rate (b), electron density (c), and inclination angle (d). In each panel, we vary one parameter while the others remain fixed. The parameter values were chosen simply to demonstrate the qualitative behavior of the model.

\begin{figure}[ht]
    \centering
    \includegraphics[width=\linewidth]{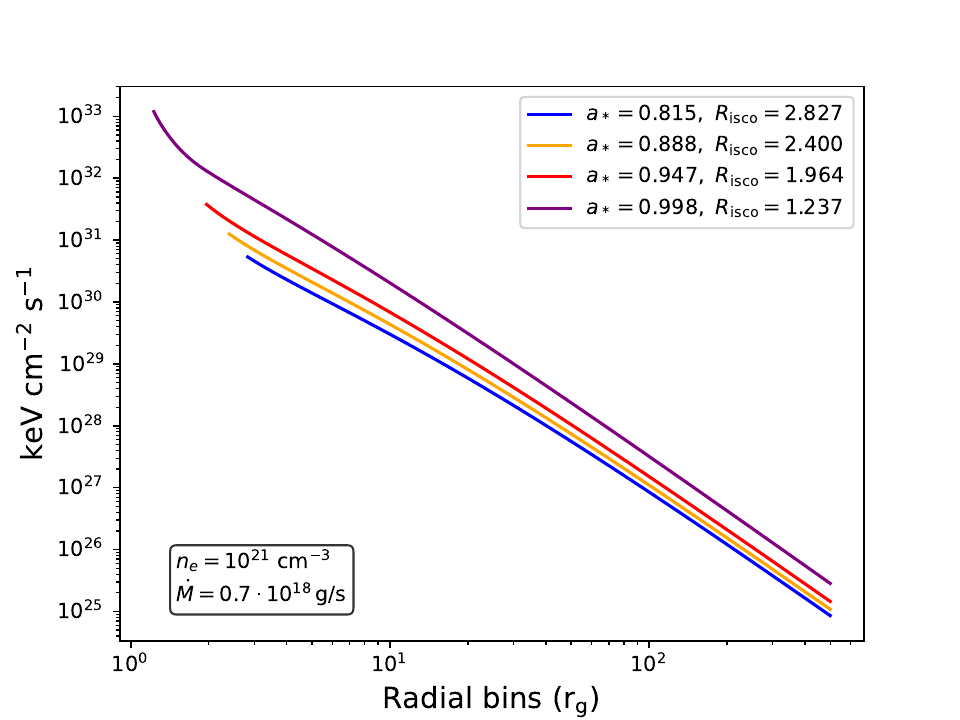}
    \caption{Returning radiation flux profile in the rest frame of the disk for a $10{\rm \, M_\odot}$ black hole and for different values of the spin parameter.}
    \label{fig:ret_rad_profile}
\end{figure}

The spin significantly shapes the reflection spectra, as shown in Fig.~\ref{fig:ziji_model_plot}a.
Furthermore, in Fig.~\ref{fig:ret_rad_profile}, we present the returning radiation flux profile in the rest frame of the disk for a $10 \,{\rm M_\odot}$ black hole at different spin values. While the overall profile shape has a strong resemblance to those reported by \citet{Wilkins-2020} and \citet{Dauser-2022}, a key difference lies in the quantity taken along the y-axis: they plotted the incident photon flux, whereas the present work computed the incident energy flux in each radial bin. An increase in spin leads to higher flux returning to each radial bin of the disk, in agreement with the enhanced flux observed at higher spins in Fig.~\ref{fig:ziji_model_plot}a, because $R_{\rm isco}$ moves closer to the black hole. Figure~\ref{fig:ziji_model_plot}b shows that increasing the mass-accretion rate primarily raises the overall flux without significantly affecting the reflection features, while panel (d) illustrates that higher inclinations shift the spectrum to high energies and "wash out" the emission lines in the soft X-rays due to stronger relativistic effects. The effect of varying the electron density, shown in panel (c), is discussed earlier in the paper. The reflection spectra in all panels share a common feature: in the absence of a corona, the Compton hump is weak or even absent, which might be a signature of sources in the high--soft state. A similar result was also reported in previous studies \citep[e.g.,][]{Mirzaev-2024a, Connors-2021}.

Finally, we generated two separate additive XSPEC table models for the relativistic spectra: one for the thermal emission (\texttt{zijiBB}) and one for the reflection component (\texttt{zijiRefl}). By combining these two files, we constructed a single model that incorporates both the reflection and blackbody components, namely \texttt{zijiRetRad}. 
All table models are available on GitHub\footnote{\href{https://github.com/kkourmpe/zijiRetRad}{\textcolor{magenta}{https://github.com/kkourmpe/zijiRetRad}}}.

We note that the current model involves a number of limitations and simplifications. These include: (1) the iron abundance is fixed at the solar value; (2) the disk is assumed to have a constant density and a constant hardening factor; and (3) the corona is not included in the calculation. These limitations arise because this type of returning radiation calculation, which includes proper radiative transfer at each location on the disk, is computationally expensive. The model should be used to assess whether returning radiation can provide an alternative interpretation of the data rather than to confirm the existence of returning radiation effects based solely on the fit. The parameter estimates obtained from this model should therefore be interpreted as purely statistical uncertainties under the assumptions adopted in the model.

\begin{table*}[ht]
\caption{Parameter grids used for the generation of the \texttt{zijiRetRad} table model.}
\centering
\renewcommand{\arraystretch}{1.5}
\begin{tabular*}{\linewidth}{@{\extracolsep{\fill}}cc}
\hline
\hline
Model Parameter & Values \\
\hline
Electron density: $n_{\rm e} \, (\mathrm{cm^{-3}})$ & $10^{18}$, $10^{19}$, $10^{20}$, $10^{21}$, $10^{22}$ \\
\hline
Spin parameter: $a_*$ & 
\begin{tabular}{l}
0.8146, 0.8346, 0.8534,
0.8710, 0.8876, 0.9035, \\ 0.9187, 0.9332, 0.9471, 0.9605, 0.9735, 0.9861, 0.9982
\end{tabular} \\
\hline
Black hole mass: $M_{ \rm BH} \, (\mathrm{M_\odot})$ & 10\\
\hline
Mass accretion rate: $\dot{M}_{\rm BH} \, \mathrm{(10^{18} \, g/s)}$ & 0.14, 0.42, 0.70, 0.84, 0.98, 1.12, 1.26, 1.40, 1.68, 2.1  \\
\hline
Inclination angle: $i \, (^\circ)$ &
\begin{tabular}{l}
30.00, 34.85, 38.87, 42.60, 
45.00, 49.50, 52.75, \\ 55.92, 59.02, 62.08, 65.12, 
 68.15, 71.21, 75.00
\end{tabular} \\
\hline
\hline
\end{tabular*}
\tablefoot{Both blackbody and reflection models share the same parameter values.}
\label{tab:xspec_parameters}
\end{table*}

\section{Spectral analysis of 4U 1630--47}\label{sec:analys}

We proceeded by testing our custom table models on the black-hole X-ray binary 4U 1630–47, using three simultaneous NuSTAR and NICER observations taken while the source was in the soft state \citep[see Fig.~2 in][]{Das-2025}. The observations are listed in Table~\ref{tab:obs}. We selected 4U 1630–47 because \cite{Connors-2021} showed that, in the soft state, the disk returning radiation is likely responsible for the reflection features.

The NuSTAR data were processed using NuSTARDAS v2.1.1 along with the NuSTAR CALDB v20230307. Cleaned event files for both FPMA and FPMB were produced using the \texttt{nupipeline} tool. Source spectra were extracted from a circular aperture of $150''$ radius centered on the source, while background spectra were taken from a source-free circular aperture of $180''$ radius. The spectra and associated data products were generated via the \texttt{nuproducts} tool. The FPMA and FPMB spectra are grouped to have a minimum count of 30 photons per bin. 

Regarding the NICER data, the \texttt{nicerl2} task was initially applied to process them using CALDB version 20221001. We then employed the \texttt{nicerl3-spect} tool to extract the source spectra, while the standard background model is used to produce the corresponding background files. We checked the power spectral density and found no signature of quasi-periodic oscillations. The spectra were subsequently rebinned following the optimal binning scheme of \cite{Kaastra-2016}, with an additional requirement of at least 30 counts per energy bin. 

The data reduction process is identical to \cite{Das-2025}. The NICER and NuSTAR data were fitted in the 1--10 keV and 4--20 keV bands, respectively. The combined spectral fitting was performed with XSPEC v12.15.0 \citep{Arnaud-1996}, using $\chi^2$ statistics to determine the best-fit values and parameter uncertainties.

\begin{table}[ht]
\caption{NICER and NuSTAR observations used in this work.}
\renewcommand{\arraystretch}{1.5}
\centering
\begin{tabular}{ccc}
\hline
\hline
 Observation ID & Time & Exposure (ks) \\
\hline
\multicolumn{3}{c}{NICER} \\
\hline
N1: 5501010104 & 2022-08-25 01:42:36 & 3.9 \\
N2: 5501010108 & 2022-08-29 06:43:00 & 3.0 \\
N3: 5501010111 & 2022-09-01 01:27:00 & 2.6 \\
\hline
\multicolumn{3}{c}{NuSTAR} \\
\hline
n1: 80802313002 & 2022-08-25 04:31:09 & 16.3 \\
n2: 80802313004 & 2022-08-29 08:16:13 & 13.2 \\
n3: 80802313006 & 2022-09-01 05:31:12 & 14.8 \\
\hline
\hline
\end{tabular}
\tablefoot{Observation 1 is composed of N1 and n1, with the same correspondence applied to the other observations.}
\label{tab:obs}
\end{table}

\subsection{Model setup}

\begin{figure}[ht]
    \centering
    \includegraphics[width=\linewidth]{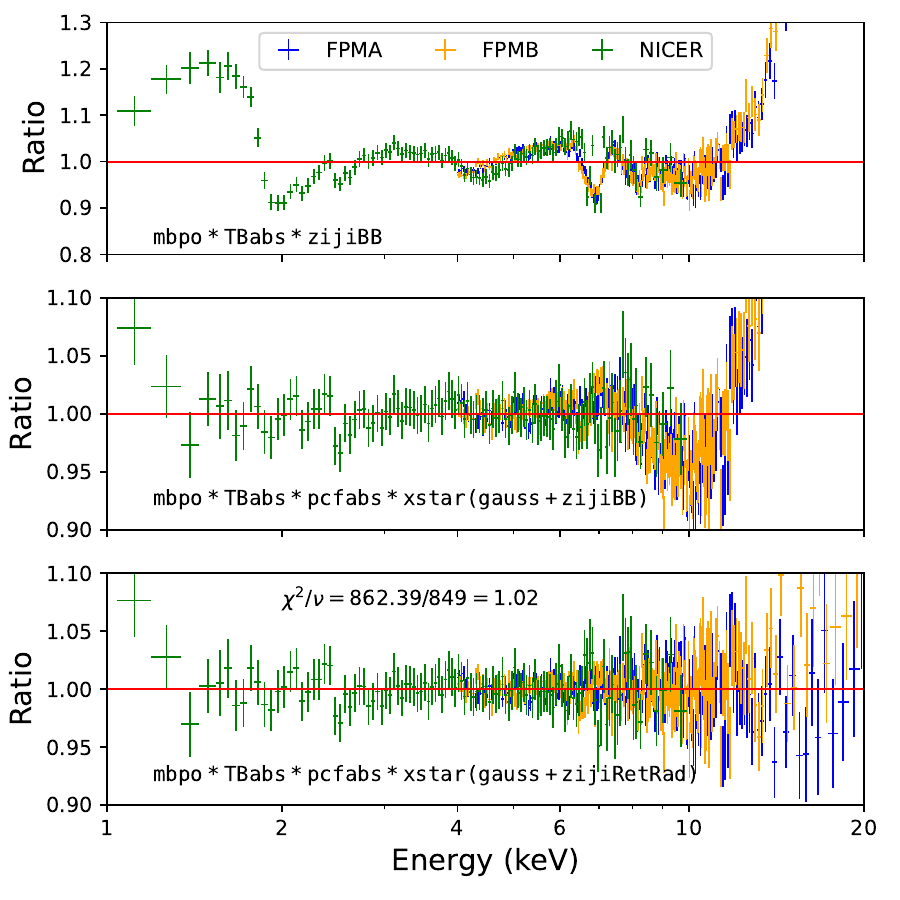}
    \caption{Ratio plots for the different models utilized during model build-up using \texttt{zijiRetRad} for Obs. 1. The respective models are shown in the bottom part of each panel. The bottom panel shows the best-fit model alongside the reduced $\chi^2$ obtained. Blue, orange, and green correspond to FPMA, FPMB, and NICER data, respectively.}
    \label{fig:ratios}
\end{figure}

To motivate the model construction using \texttt{zijiRetRad}, we present the process step by step, justifying the inclusion of each component for Obs1. We fixed the black-hole mass to $10 \, \mathrm{M_\odot}$ (\citeauthorandyear{Seifina-2014}; indirect mass estimate) and begin with the baseline model (in XSPEC language)  \texttt{mbpo*TBabs*zijiBB}, where \texttt{mbpo}\footnote{The \texttt{mbpo} model multiplies the total model by a constant  (or normalization) and a broken power law, with indices $d\Gamma_1$ for $E < E_{\rm br}$ and $d\Gamma_2$ for $E > E_{\rm br}$, where $E_{\rm br}$ is the break energy. For the NICER data, all model parameters are left free. For the NuSTAR data, the constant is fixed to unity for FPMA and left free for FPMB, while all power-law indices are fixed to zero.
} \citep{Ingram-2017} accounts for cross-calibration differences between NuSTAR and NICER, \texttt{TBabs} models interstellar absorption \citep{Wilms-2000}, and \texttt{zijiBB} is the disk's blackbody table model generated with the \texttt{ziji} code. 
The performance of this model is shown in the top panel of Fig.~\ref{fig:ratios}. Several features remain unaccounted for, including an edge near 1.7 keV and prominent reflection signatures above 6 keV. In particular, the sharp rise beyond 10 keV likely marks the onset of the Compton hump, which cannot be reproduced by a pure blackbody model.

Motivated by previous work \citep{Kushwaha-2023, Das-2025}, we included a partial-covering absorber using \texttt{pcfabs}. In addition, the residual feature at $\sim$1.7 keV (Fig.~\ref{fig:ratios}, top panel), attributed to Au fluorescence from the NICER optics, was modeled with a narrow Gaussian component (\texttt{gauss}) \citep{Das-2025, Kushwaha-2023}. Significant absorption residuals also persist in the 6–7 keV range, corresponding to disk--wind features previously identified as Fe XXV and Fe XXVI absorption lines at 6.697 keV and 6.966 keV, respectively \citep{Bianchi-2005, King-2013, Pahari-2018}. These were modeled using a custom photoionization table generated with \texttt{xstar} \citep{Kallman-2001}, adopting the setup of \cite{Das-2025}, with a turbulent velocity of $1000 \,\rm{km \,s^{-1}}$ and a plasma density of $10^{12} \,\rm{cm^{-3}}$. Further details on the table construction are given in \cite{Das-2025}.

With these additions, the model performance (Fig.~\ref{fig:ratios}, middle panel) improves significantly at low energies, although residuals remain above 6 keV. Finally, we included reflection from thermal returning radiation via \texttt{zijiRetRad}, yielding the best-fit model:
\begin{center}
\texttt{mbpo*TBabs*pcfabs*xstar(gauss+zijiRetRad)}.
\end{center}

The resulting fit (Fig.~\ref{fig:ratios}, bottom panel) shows no significant systematic residuals, with a $\chi^2/\nu = 1.02$. The corresponding best-fit parameters are listed in Table~\ref{tab:bestfit} (first column). Parameter uncertainties were estimated at the 90\% confidence level using \texttt{steppar} followed by \texttt{error}.

\subsection{Results and discussion}

The fits to the three observations yield a rapidly spinning black hole $(a_* \sim 0.99)$ and a disk inclination of $i \sim 43^\circ-54^\circ$. The hydrogen column density is estimated to be $N_{\rm H} \sim 7 \cdot 10^{22} \, \mathrm{cm^{-2}}$, while the partial-covering absorber has a similar column density, with a covering fraction close to unity. The mass-accretion rate is  in the $\dot{M}_{\rm BH} \sim (0.7-1.5) \cdot 10^{18} \, {\rm g/s}$ range, corresponding to approximately $(0.13-0.25) \, \dot{M}_{\rm Edd}$. Moreover, we inferred a high electron density on the order of $n_e \sim 10^{21} \, \mathrm{cm^{-3}}$. Finally, the distance to the source inferred from our analysis is $D \sim 7.3-11.3 \, {\rm kpc}$. Overall, the parameter values obtained in our analysis are broadly consistent with those reported in the literature, and the  best-fit values are presented in Table~\ref{tab:bestfit}.

We modeled the absorption using a combination of full-covering (\texttt{TBabs}) and partial-covering (\texttt{pcfabs}) components. For covering fractions close to unity, the effective column density approaches the sum of the two absorbers, yielding values consistent with previous studies that employed a single, fully covering absorber. \cite{Kuulkers-1998}, \cite{Connors-2021}, \cite{Das-2025}, and \cite{Tomsick-1998} all reported high hydrogen column densities along the line of sight, with $N_{\rm H} \sim (4-12) \cdot 10^{22} \, \mathrm{cm^{-2}}$. \cite{Gatuzz-2019} reported a value closer to $10^{23} \, \mathrm{cm^{-2}}$, while \cite{Connors-2021} adopted $N_{\rm H} \sim 1.4 \cdot 10^{23} \, \mathrm{cm^{-2}}$. They justified this choice by pointing to the uncertainties in $N_{\rm H}$ derived from X-ray data \citep{Tomsick-1998, Tomsick-2000, Gatuzz-2019}, as well as the presence of dust scattering along the line of sight \citep{Kalemci-2018}. The mass-accretion rate is consistent with the value reported by \cite{Das-2025} $(\dot{M}_{\rm BH} \sim (0.15-0.2) \, \dot{M}_{\rm Edd})$.

A high inclination has been reported by \cite{Tomsick-1998}, \cite{Seifina-2014}, and \cite{Connors-2021}, with values in the range of $60^\circ<i<70^\circ$. However, \cite{Connors-2021} also found a lower estimate of $i \sim 37^\circ$ using \texttt{relxillNS}. Reflection modeling by \cite{King-2014} provided an estimate of $i \sim 64^\circ$. Recent analyses by \cite{Draghis-2024} and \cite{Das-2025} reported somewhat lower inclination values of $i \sim 55^\circ$, with the former exhibiting larger uncertainties; these values are closer to the ones obtained in our analysis.

The spin of the black hole has also been estimated to be high. \cite{King-2014} found $a_* = 0.985^{+0.005}_{-0.014}$, and \cite{Connors-2021} reported a similar value, although lower estimates have also appeared in the literature, for example\
$a_* = 0.857^{+0.095}_{-0.211}$ \citep{Draghis-2024},
$0.85^{+0.07}_{-0.07}$ \citep{Connors-2021}, and
$0.837^{+0.067}_{-0.156}$ \citep{Das-2025}. It is worth noting that the statistical uncertainties on the spin in our fits are not physically meaningful and underestimate the true uncertainty due to systematic effects and model assumptions. Our result should therefore be interpreted as evidence for a high-spin, returning radiation scenario rather than a precise measurement of the black-hole spin.

The high electron density we find in our analysis is consistent with the results of \cite{Connors-2021}, who reported $n_{\rm e} > 10^{20} \, {\rm cm^{-3}}$ across all spectral states. As discussed earlier (see Fig.~\ref{fig:ziji_model_plot}c), such high densities are required for the reflection features to be present. In addition, the parameters we obtain for the \texttt{xstar} and \texttt{mbpo} models are in good agreement with those reported by \cite{Das-2025} when analyzing the same observations. 

The distance inferred in our analysis is close to the values reported by \cite{Kalemci-2025} and \cite{Kalemci-2018} when modeling the dust-scattering halo $(\sim 11.5 \, {\rm kpc})$, as well as the $10 \, {\rm kpc}$ distance commonly adopted in the literature \citep[e.g.,][]{Seifina-2014, Augusteijn-2001}. However, it is important to note that the mass-accretion rate and the distance are highly degenerate parameters. This is also evident from the fits to the individual observations, where these parameters vary in a compensatory manner despite the sources being in very similar spectral states. As a result, our estimates for these quantities remain model dependent.

The model components and residuals of the three observations fitted using \texttt{zijiRetRad} are shown in Fig.~\ref{fig:mdl-delchi}. Obs. 1 provides an excellent fit to the data, with no significant residuals. In contrast, the fits for Obs. 2 and Obs. 3 progressively worsen, with a narrow Compton hump still visible in the residuals of the latter. This could point to limitations in describing the observed Compton hump with disk self-irradiation alone, indicating the presence of an additional coronal component contributing to the disk illumination.

Although our results are broadly consistent with previous studies, the $\sim10^\circ$ variation in the inferred inclination suggests that the model cannot adequately describe all three observations with a single source geometry. We therefore performed a simultaneous fit to all three observations, tying $D$, $i$, $a_*$, and $N_{\rm H}$ across the datasets, as well as the instrumental cross-calibration parameters.
The simultaneous fit yields $a_*\sim0.99$, $i \sim 47.5^\circ$, and $D \sim 9.65 \,\rm{kpc}$, with a good fit quality of $\chi^2/\nu=2847.90/2500=1.14$. These results are consistent with a rapidly spinning black hole. The simultaneous fit also gives lower electron density values for Obs. 1 and 2 than in the untied fits, with densities on the order of $10^{20} \,\rm{cm^{-3}}$. Finally, the mass-accretion rates show smaller variations among the three observations in the simultaneous fit and are consistent with the values reported by \cite{Das-2025}, whose authors analyzed the same dataset using tied parameters. The complete set of parameters obtained from this fit is presented in Table~\ref{tab:bestfit_combined}.

\begin{table*}[ht]
\caption{Best-fit values obtained using \texttt{zijiRetRad}.}
    \centering
    \renewcommand{\arraystretch}{1.5}
    \begin{tabular*}{\linewidth}{@{\extracolsep{\fill}}cccccc}
    \hline
    \hline
    Component          & Parameter     & Obs. 1 & Obs. 2 & Obs. 3  \\
    \hline
    \texttt{TBabs}     & $N_{\rm H} \, (10^{22} \, \mathrm{cm^{-2}})$  & $6.8_{-0.5}^{+0.4}$  & $7.1^{+0.4}_{-0.5}$ & $7.8_{-0.4}^{+0.3}$  \\
    \hline
    \texttt{pcfabs}     & $N_{\rm H} \, (10^{22} \, \mathrm{cm^{-2}})$  & $7.3_{-0.3}^{+0.3}$  & $7.4^{+0.3}_{-0.2}$  & $7.0_{-0.2}^{+0.2}$ \\
                       & $CvrFract$                             & $0.93_{-0.02}^{+0.02}$  & $0.91^{+0.02}_{-0.03}$ & $0.870_{-0.015}^{+0.023}$    \\
    \hline
    \texttt{xstar}     & $column \, (10^{23})$                  & $1.27_{-0.14}^{+0.14}$  & $1.35^{+0.14}_{-0.27}$ & $1.08_{-0.07}^{+0.05}$   \\
                       & $rlog\xi$                             & $4.11_{-0.08}^{+0.11}$  & $4.09^{+0.10}_{-0.17}$ & $3.90_{-0.02}^{+0.03}$     \\
    \hline
    \texttt{gaussian} & $LineE \, \rm{(keV)}$ & $1.71_{-0.02}^{+0.02}$      & $1.730_{-0.016}^{+0.014}$      & $1.72_{-0.02}^{+0.02}$ \\
                      & $sigma \, (10^{-2})$ & $7.1_{-1.9}^{+2.2}$            &  $5.3_{-1.8}^{+1.9}$      &  $6.7_{-1.6}^{+2.0}$\\
                      & $norm \, (10^{-2})$  & $11.6_{-3.0}^{+4.0}$   &  $9.4_{-2.2}^{+3.0}$   &  $9.6_{-2.0}^{+2.7}$ \\
    \hline
    \texttt{zijiRetRad} & $n_{\rm e} \, \mathrm{(10^{21} \, cm^{-3})}$ & $0.99_{-0.46}^{+0.94}$    & $1.0^{+0.9}_{-0.4}$ & $0.7_{-0.3}^{+0.4}$ \\
                       & $a_*$                                 & $0.9878_{-0.0003}^{+0.0004}$  & $0.9869^{+0.0001}_{-0.0007}$ & $0.9860_{-0.0014}^{+0.0002}$    \\
                       & $M_{\rm BH} \, \mathrm{(M_\odot)}$           &   $10^*$ &   $10^*$   &   $10^*$   \\
                       & $\dot{M}_{\rm BH} \, \mathrm{(10^{18} \, g/s)}$ & $0.82_{-0.14}^{+0.03}$    & $1.17^{+0.20}_{-0.10}$ & $1.47_{-0.06}^{+0.03}$ \\
                       & $i \, (^\circ)$ & $52.8_{-0.5}^{+1.0}$   & $47.4^{+1.1}_{-2.4}$   & $43.4_{-0.3}^{+0.8}$      \\
                       & $D \, {(\rm kpc)}$ & $8.1_{-0.8}^{+0.2}$   &   $9.8^{+1.3}_{-0.4}$  & $11.0_{-0.2}^{+0.2}$
                       \\
                       & $f_{\rm col}$ & $1.7^*$ & $1.7^*$ & $1.7^*$ \\
                       & $A_{\rm Fe}$ & $1.0^*$ & $1.0^*$ & $1.0^*$ \\
                       
    \hline
     \texttt{mbpo} & $C_{\rm FPMB}$  &   $0.987_{-0.002}^{+0.002}$   & $0.992^{+0.002}_{-0.002}$ & $0.9890_{-0.0014}^{+0.0015}$    \\
      &  $C_{\rm NICER}$   &     $0.990_{-0.013}^{+0.022}$         &   $1.047_{-0.016}^{+0.012}$       &  $1.046_{-0.028}^{+0.013}$ \\
      & $d\Gamma_1$         &     $0.21_{-0.03}^{+0.02}$        &   $0.21_{-0.02}^{+0.02}$       &  $0.210_{-0.014}^{+0.013}$  \\
      & $d\Gamma_2$         &     $0.07_{-0.09}^{+0.05}$        &   $-0.08_{-0.18}^{+0.18}$       &  $0.014_{-0.135}^{+0.124}$  \\
      & $E_{\rm br}$        &      $6.0_{-0.8}^{+0.8}$       &    $7.6_{-1.1}^{+0.4}$      &  $7.36_{-1.03}^{+0.44}$   \\  
      \hline
    $\chi^2/\nu$        &                                      & $862.39/849=1.02$        & $913.30/818=1.12$   & $947.23/811=1.17$   \\
    \hline                                                    
    \hline
    \end{tabular*}
    \tablefoot{The asterisk ($*$) denotes that the parameter is fixed at this value.}
    \label{tab:bestfit}
\end{table*}

\begin{table*}[ht]
\caption{Best-fit values obtained from the simultaneous fit using \texttt{zijiRetRad}.}
    \centering
    \renewcommand{\arraystretch}{1.5}
    \begin{tabular*}{\linewidth}{@{\extracolsep{\fill}}cccccc}
    \hline
    \hline
    Component          & Parameter     & Obs. 1 & Obs. 2 & Obs. 3  \\
    \hline
    \texttt{TBabs}     & $N_{\rm H} \, (10^{22} \, \mathrm{cm^{-2}})$  & $7.36_{-0.21}^{+0.19}$  & tied & tied  \\
    \hline
    \texttt{pcfabs}     & $N_{\rm H} \, (10^{22} \, \mathrm{cm^{-2}})$  & $7.35_{-0.06}^{+0.13}$  & $7.32_{-0.07}^{+0.12}$  & $6.98_{-0.14}^{+0.13}$ \\
                       & $CvrFract$                             & $0.894_{-0.015}^{+0.008}$  & $0.892_{-0.015}^{+0.006}$ & $0.901_{-0.014}^{+0.008}$    \\
    \hline
    \texttt{xstar}     & $column \, (10^{23})$                  & $1.15_{-0.16}^{+0.14}$  & $1.38_{-0.22}^{+0.13}$ & $1.44_{-0.10}^{+0.10}$   \\
                       & $rlog\xi$                             & $4.05_{-0.11}^{+0.08}$  & $4.10_{-0.14}^{+0.09}$ & $4.12_{-0.07}^{+0.07}$     \\
    \hline
    \texttt{gaussian} & $LineE \, \rm{(keV)}$ & $1.724_{-0.009}^{+0.009}$      & tied       & tied  \\
                      & $sigma \, (10^{-2})$
                      & $6.0_{-1.0}^{+1.1}$            &  tied      &  tied\\
                      & $norm \, (10^{-2})$  & $8.7_{-1.7}^{+1.9}$   &  $9.6_{-1.8}^{+2.0}$   &  $10.0_{-2.0}^{+2.2}$ \\
    \hline
    \texttt{zijiRetRad} & $n_{\rm e} \, \mathrm{(10^{21} \, cm^{-3})}$ & $0.097_{-0.009}^{+0.037}$    & $0.090_{-0.009}^{+0.008}$ & $1.00_{-0.04}^{+0.63}$ \\
                       & $a_*$                                 & $0.98659_{-0.00007}^{+0.00005}$  & tied & tied    \\
                       & $M_{\rm BH} \, \mathrm{(M_\odot)}$           &   $10^*$ &   $10^*$   &   $10^*$   \\
                       & $\dot{M}_{\rm BH} \, \mathrm{(10^{18} \, g/s)}$ & $1.153_{-0.015}^{+0.010}$    & $1.112_{-0.008}^{+0.007}$ & $1.124_{-0.028}^{+0.013}$ \\
                       & $i \, (^\circ)$
                       & $47.53_{-0.36}^{+0.17}$   & tied   & tied     \\
                       & $D \, {(\rm kpc)}$ & $9.65_{-0.11}^{+0.05}$   &   tied  & tied
                       \\
                       & $f_{\rm col}$ & $1.7^*$ & $1.7^*$ & $1.7^*$ \\
                       & $A_{\rm Fe}$ & $1.0^*$ & $1.0^*$ & $1.0^*$ \\
                       
    \hline
     \texttt{mbpo} & $C_{\rm FPMB}$  &   $0.9892_{-0.0011}^{+0.0011}$   & tied & tied    \\
      &  $C_{\rm NICER}$   &     $1.016_{-0.008}^{+0.008}$         &   tied       &  tied \\
      & $d\Gamma_1$         &     $0.204_{-0.011}^{+0.010}$        &   tied       &  tied  \\
      & $d\Gamma_2$         &     $0.06_{-0.05}^{+0.04}$        &   tied       &  tied  \\
      & $E_{\rm br}$        &      $6.6_{-0.3}^{+0.3}$       &    tied     &  tied   \\  
      \hline
    $\chi^2/\nu$    &  & & $ 2847.90/2500=1.14$ &     \\
    \hline                                                    
    \hline
    \end{tabular*}
    \label{tab:bestfit_combined}
\end{table*}

\subsection{Comparison with \texttt{relxillNS}}  \label{sec:zj_vs_NS}

To put our results into context, we compare \texttt{zijiRetRad} with  \texttt{relxillNS} \citep{Garcia-2022}, a model widely applied for returning radiation \citep[e.g.,][]{Connors-2020,Connors-2021, Rout-2023}. We stress, however, that the two models are based on different physical assumptions and are therefore not directly equivalent. In particular, \texttt{relxillNS} does not model self-irradiation of the disk using its intrinsic emission; instead, it assumes illumination by a single-temperature blackbody, mimicking the effect of returning radiation. In addition, the models differ in their treatment of radiative transfer, with \texttt{zijiRetRad} relying on the \texttt{reflionx} \citep{Ross_Fabian-2005}, while \texttt{relxillNS} uses the more recent \texttt{xillverNS} \citep{Garcia-2022}.

\begin{figure*}[ht]
    \centering
    \includegraphics[width=\linewidth]{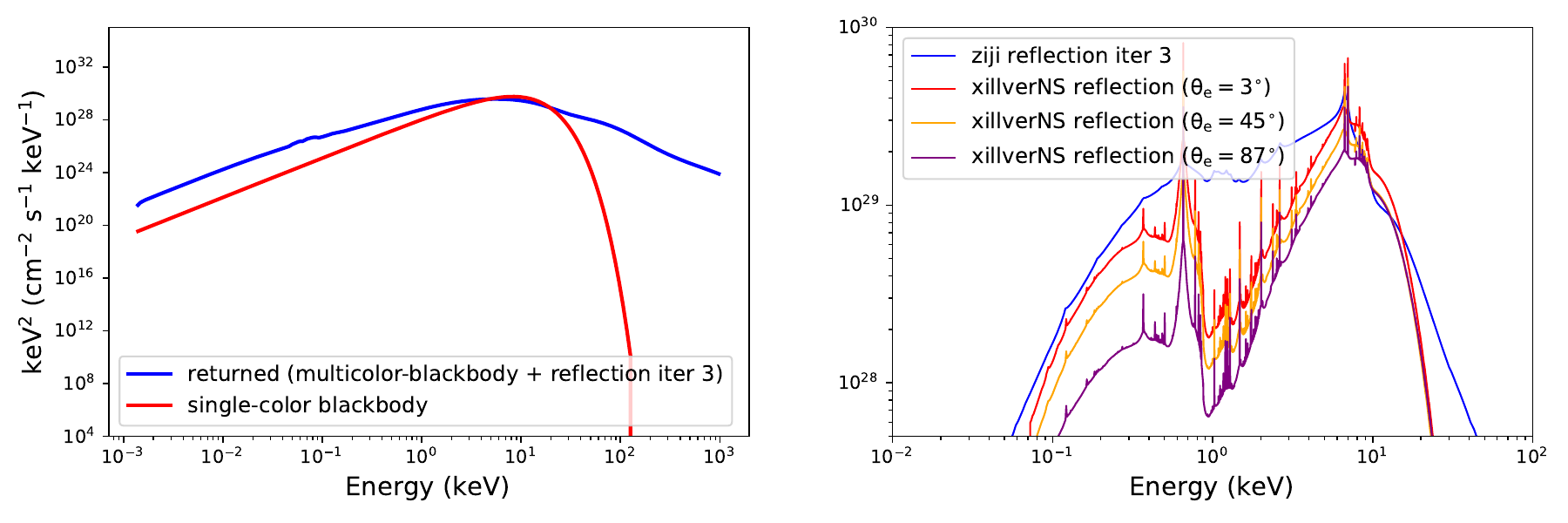}
    \caption{\textit{Left}: Accretion-disk incident spectra. Single-color blackbody assumed in \texttt{relxillNS} (red curve), and the total incident spectra of the third iteration given by the sum of the returning thermal emission and the returning reflection component, as calculated by \texttt{zijiRetRad} (blue curve). \textit{Right}: Comparison of disk reflection spectra. The red, orange, and purple curves show the \texttt{xillverNS} reflection generated by the illumination of the closest single-color blackbody, respectively, for emission angles of $\theta_{\rm e} = 3^{\circ}, 45^{\circ}$, and $87^{\circ}$. The blue curve shows the third-iteration \texttt{zijiRetRad} reflection produced by the illumination of the returning multicolored blackbody plus the returning reflection. We also used the following parameters: $M_{\rm BH} = 10 \,\rm{M_{\odot}}$, $a_*= 0.998$, $\dot{M}_{\rm BH} = 8.2\cdot 10^{17}~\rm{g/s}$, $kT_{\rm bb} = 2.1~\rm{keV}$, $A_{\rm Fe} = 1.0$, $log(n_{e}) = 19~{\rm cm^{-3}}$, and  $log\xi = 3.2~{\rm erg \cdot cm/s}$. The \texttt{xillverNS} model flux was normalized to match the iron line flux of the \texttt{zijiRetRad} model.}
    \label{fig: incident_and_reflection}
\end{figure*}

The left panel of Fig.~\ref{fig: incident_and_reflection} compares the incident spectra computed in \texttt{zijiRetRad} (blue curve) and assumed by  \texttt{xillverNS} (red curve). The former was calculated self-consistently by summing the returning thermal spectrum and the third-iteration returning reflection spectrum, whereas the latter was modeled as a single-temperature blackbody. In order to make a fair comparison of the incident and reflection spectra, we proceeded as follows. As \texttt{zijiRetRad} assumes an accretion disk of constant density and computes variable ionization across radial bins due to the varying incident flux at each bin, we selected a radial bin whose ionization value falls within the grid covered by \texttt{xillverNS}, enabling a direct comparison. From this fixed radial bin, we extracted the \texttt{zijiRetRad} incident spectra and the corresponding reflection spectra. For Fig.~\ref{fig: incident_and_reflection}, we chose the radial bin $r = 16.72~r_{\rm g}$, corresponding to $log\xi = 3.2\ \rm{erg \cdot cm/s}$. To calculate the effective temperature for the closest single-color blackbody spectrum, we first integrated the \texttt{zijiRetRad} incident spectra over energy, and subsequently applied Wien's displacement law to derive the corresponding temperature. The peak of the resulting spectrum in $\nu F_{\nu}$ units lies around 8~keV. In this specific example, below 8~keV the two spectra show moderate differences, with \texttt{zijiRetRad} predicting a higher flux; this discrepancy increases toward lower energies. This difference is due to both returning blackbody and returning reflection photons. Above 8~keV, however, the difference becomes pronounced, with \texttt{zijiRetRad} predicting a significantly higher flux. This excess is mainly due to a substantial fraction of reflection photons being blueshifted as they return to other parts of the disk, producing a high-energy tail. In the right panel of Fig.~\ref{fig: incident_and_reflection}, we compare the rest-frame\footnote{We omit a comparison of the two reflection spectra in the distant observer frame because the calculations performed by \texttt{zijiRetRad} and \texttt{relxillNS} to project the spectra into the distant frame are essentially identical.} spectra produced by \texttt{zijiRetRad} (blue) and \texttt{xillverNS} (red, yellow, and purple) for the two corresponding incident spectra. We set the emission angle in \texttt{xillverNS} to three distinct values: $3^{\circ}, 45^{\circ}$, and $87^{\circ}$, representing face-on, intermediate, and edge-on viewing angles, respectively. We kept the original normalization native to the model for \texttt{zijiRetRad} and normalized the \texttt{xillverNS} spectra to match the flux of the \texttt{zijiRetRad} spectrum at the iron K$\alpha$ line. The two reflection spectra exhibit significant differences. Overall, \texttt{xillverNS} produces significantly more emission lines than \texttt{reflionx}. This is due to a superior atomic database considered in \texttt{xillverNS}. Above the Fe $K\alpha$ line, the reflection spectra from both models exhibit a moderate Compton hump; however, \texttt{reflionx} comparatively produces more hard photons at higher energies. This makes sense because \texttt{reflionx's} incident spectrum contains a higher flux of hard photons. Below the Fe $K\alpha$ line, the difference between the spectra becomes more pronounced; \texttt{xillverNS} exhibits higher absorption \citep[see][]{Garcia-2011}. Given these differences, the comparison presented here is intended to illustrate how distinct modeling assumptions map onto the local and finally onto observed spectral shape, rather than to provide a direct one-to-one test between the models.

 Since \texttt{relxillNS} has been extensively studied in the literature, we do not present its setup in detail here. We fitted the same observational data with \texttt{relxillNS}, and the resulting best-fit model is
\begin{center}
    \texttt{mbpo*TBabs*pcfabs*xstar* \\ (gauss+nthComp+kerrbb+relxillNS)},
\end{center}
where \texttt{nthComp} accounts for the Comptonized emission (assuming a multicolored disk blackbody-seed spectrum; $inp\_type = 1$) and \texttt{kerrbb} for the multicolored blackbody. We fixed the power-law index, $\Gamma = 2.5$, and the electron temperature, $kT_{\rm e} = 100 \, {\rm keV}$, of \texttt{nthComp}, leaving only the normalization as a free parameter. Moreover, the distance is fixed at $D = 10 \, {\rm kpc}$ \citep{Seifina-2014,Augusteijn-2001}, since it cannot be simultaneously constrained with the mass-accretion rate within this model. The mass is fixed at $10 \, {\rm M_\odot}$ \citep{Seifina-2014} and the iron abundance at the solar value $(A_{\rm Fe}=1)$, ensuring consistency with the prior analysis using \texttt{zijiRetRad}. The setup of the \texttt{mbpo} model is the same as in the fitting using \texttt{zijiRetRad}. For reference, \texttt{zijiBB} and \texttt{kerrbb} are essentially equivalent.

The best-fit values obtained with \texttt{relxillNS} are listed in Table~\ref{tab:bestfit_NS}. They are broadly consistent with earlier studies and with the results from \texttt{zijiRetRad}, although some differences are present. Overall, we obtain a high black-hole spin $(a_*\sim 0.996)$, an inclination of $i\sim 57^\circ$, and a mass-accretion rate of $\dot{M}_{\rm BH}\sim 0.6 \cdot 10^{18} \, \rm{g/s}$. The main goal of this comparison is to examine the resulting spectral components, while a detailed discussion of all parameter values is beyond the scope of this work, given that \texttt{relxillNS} has already been applied to this observation in previous studies \citep[see, e.g.,][]{Das-2025}. Although \texttt{relxillNS} is not meant to capture the same physics in a fully self-consistent way, it provides a statistically better fit to the data, which reflects the flexibility of the model.

\begin{figure*}
    \centering
    \includegraphics[width=\linewidth]{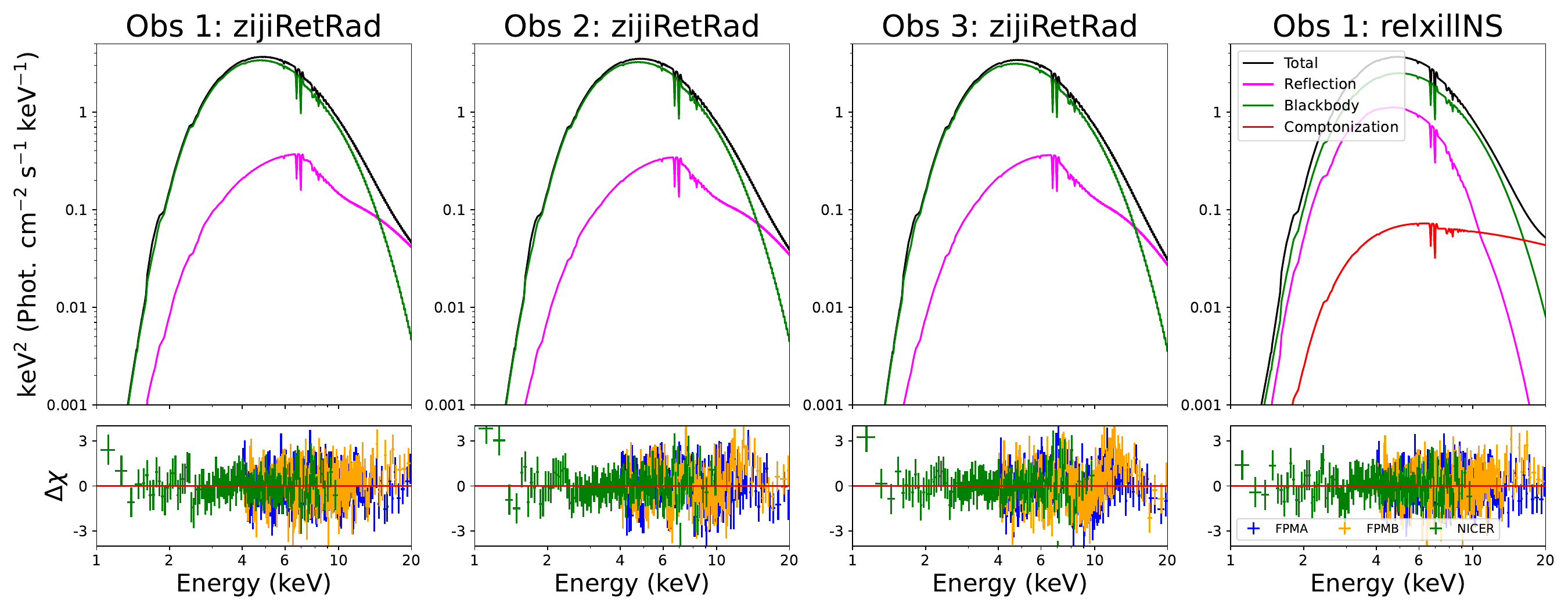}
    \caption{Model components (\textit{top row}) and $\Delta \chi={\rm (data-model)/error}$ (\textit{bottom row}) for the best-fit models obtained with \texttt{zijiRetRad} (Obs. 1-3) and \texttt{relxillNS} (Obs 1). The green, magenta, red, and black curves correspond to the blackbody, reflection, Comptonization, and total spectra, respectively. In the \texttt{zijiRetRad} (\texttt{relxillNS}) panels, the reflection and blackbody components correspond to \texttt{zijiRefl} (\texttt{relxillNS}) and \texttt{zijiBB} (\texttt{kerrbb}), while the Comptonization component is present only in the \texttt{relxillNS} panel. In the \texttt{zijiRetRad} panels the model is \texttt{mbpo * TBabs * pcfabs * xstar (gauss + zijiRetRad)}, while for the \texttt{relxillNS} one the model reads as follows: \texttt{mbpo * TBabs * pcfabs * xstar (gauss + nthComp + kerrbb + relxillNS)}.}
    \label{fig:mdl-delchi}
\end{figure*}

\begin{figure*}[ht]
    \centering
    \includegraphics[width=\linewidth]{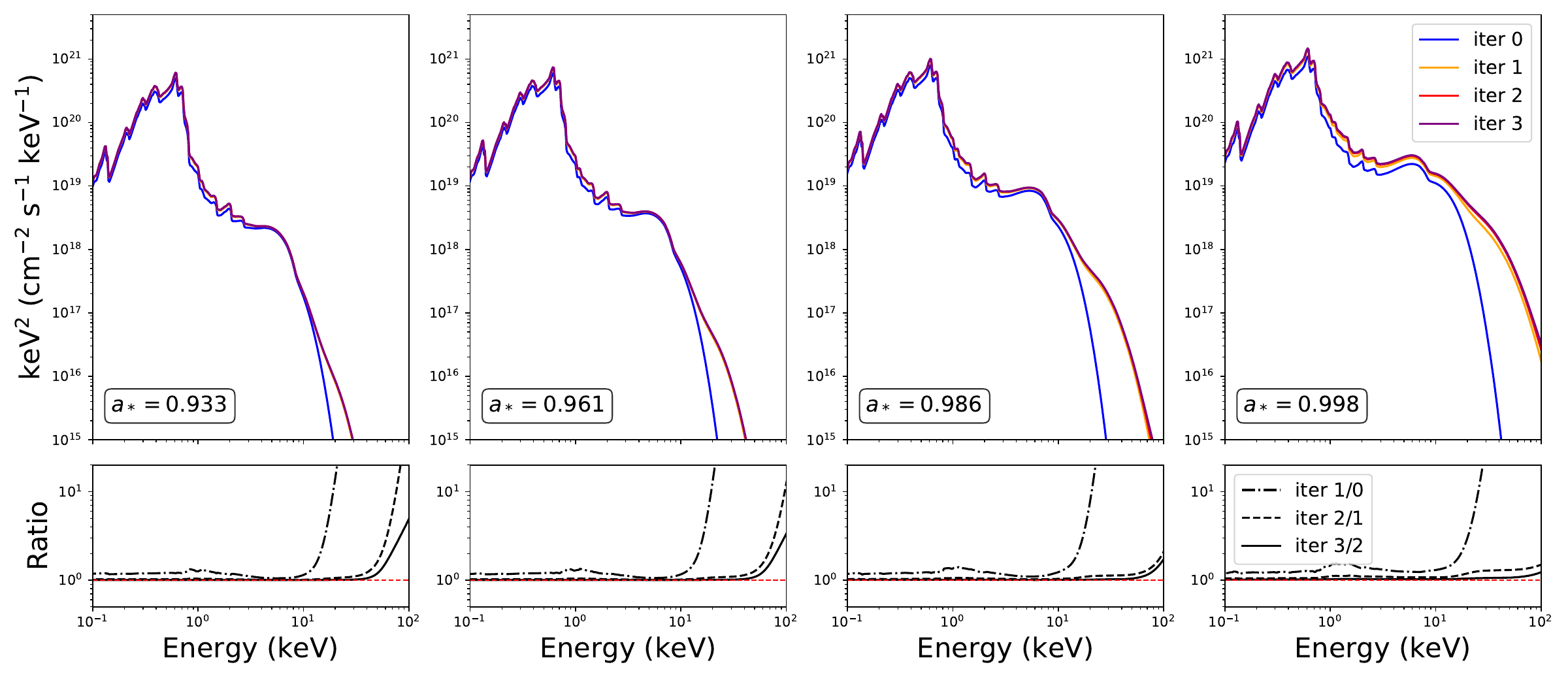}
    \caption{\textit{Top panels}: Relativistic reflection spectra generated with the \texttt{ziji} code for different numbers of iterative reflections (0–3) and spin values. \textit{Bottom panels}: Ratio between the flux of each successive iteration. Model parameters were randomly selected to illustrate the qualitative behavior of the spectrum as the number of iterations increases (in all panels: $M_{\rm BH}=10 \, {\rm M_\odot}, \dot{M}_{\rm BH}=0.3 \cdot 10^{18} \, {\rm g/s}, n_{\rm e} = \mathrm{10^{22} \, cm^{-3}}, i=45^\circ$). Iteration 0 represents the reflection spectra produced solely by the direct returning radiation of the thermal component, while iterations 1–3 show the reflection spectra that also include the returning radiation of the reflection component for the corresponding number of iterations.}
    \label{fig:it_model}
\end{figure*}

The model components and residuals are shown in the right panels of Fig.~\ref{fig:mdl-delchi}. The most notable difference between the two models is that the \texttt{zijiRetRad} reflection spectrum is significantly harder at high energies relative to \texttt{relxillNS}. Consequently, only the latter one requires an additional weak coronal component to produce the observed high-energy tail. In the case of \texttt{zijiRetRad}, the hard excess arises from the iterative reflection included in the model, which is absent in \texttt{relxillNS}. Figure~\ref{fig:it_model} shows the relativistic reflection spectra for different numbers of iterations (0–3) and for varying black-hole spins. The high-energy tail becomes progressively harder with increasing iteration number and black-hole spin. As the disk $(R_{\rm isco})$ moves closer to the black hole, gravitational light bending and Doppler blueshift increase the flux reaching the disk from its innermost regions, resulting in a harder high-energy spectrum for the iterative reflection. Interestingly, without accounting for a coronal component, producing a high-energy tail in the reflection spectrum requires pushing the spin parameter to high values. As a result, neglecting the corona could lead to overestimating the inferred spin in some cases, when such tails are present in the data. However, in our case, we verified that adding a coronal component alongside \texttt{zijiRetRad} in XSPEC does not improve the quality of the fit.

\begin{table}[ht]
\caption{Same as Table~\ref{tab:bestfit} for Obs. 1 but for \texttt{relxillNS}.}
    \centering
    \renewcommand{\arraystretch}{1.5}
   \begin{tabular*}{\columnwidth}{@{\extracolsep{\fill}}ccc}
    \hline
    \hline
    Component          & Parameter                             & Obs. 1  \\
    \hline
    \texttt{TBabs}     & $N_{\rm H} \, (10^{22} \, \mathrm{cm^{-2}})$  & $4.60_{-0.62}^{+0.02}$   \\
    \hline
    \texttt{pcfabs}     & $N_{\rm H} \, (10^{22} \, \mathrm{cm^{-2}})$  & $8.0_{-0.5}^{+0.5}$   \\
                       & $CvrFract$                             & $0.985_{-0.024}^{+0.015}$    \\
    \hline
    \texttt{xstar}     & $column \, (10^{23})$                  & $1.23_{-0.04}^{+0.04}$   \\
                       & $rlog\xi$                             & $4.12_{-0.02}^{+0.02}$ \\
    \hline
    \texttt{gaussian} & $LineE \, \rm{(keV)}$ & $1.675_{-0.011}^{+0.010}$  \\
                      & $sigma \, (10^{-2})$ & $10.8_{-0.9}^{+1.0}$      \\
                      & $norm \, (10^{-2})$  & $19.1_{-1.4}^{+1.4}$   \\
    \hline
    \texttt{nthComp}  & $Gamma$   & $2.5^*$\\
                   & $kT_{\rm e} \, {\rm (keV)}$ &   $100^*$\\
                   & $norm \, (10^{-2})$ &  $4.70^{+0.11}_{-0.10}$ \\
    \hline
    \texttt{kerrbb} &     $a_*$                     &  $0.99619^{+0.00005}_{-0.00534}$                \\
                    &     $i \, (^\circ)$      &  $57.3^{+1.1}_{-1.1}$                 \\
                    &     $\dot{M}_{\rm BH} \, \mathrm{(10^{18} \, g/s)}$ & $0.60^{+0.23}_{-0.23}$ \\
                    &     $M_{\rm BH} \, {\rm (M_\odot)}$     & $10^*$  \\
                    &     $D \, {\rm (kpc)}$          & $10^*$\\
    \hline
    \texttt{relxillNS} & $kT_{{\rm bb}} \, {\rm (keV)}$ &   $0.898^{+0.002}_{-0.002}$    \\
                       & $log\xi$              &    $2.560^{+0.010}_{-0.010}$     \\
                       & $log(n_{\rm e}) \, {\rm (cm^{-3})}$           & $17.50^{+0.10}_{-0.10}$       \\
                       & $A_{\rm Fe}$ &   $1.0^*$   \\
                       & $norm \, (10^{-2})$ &    $1.7^{+0.3}_{-0.3}$                  \\
    \hline
    \texttt{mbpo} & $C_{\rm FPMB}$  &   $0.987_{-0.002}^{+0.002}$   \\
      &  $C_{\rm NICER}$   &     $0.991_{-0.003}^{+0.003}$        \\
      & $d\Gamma_1$         &     $0.201_{-0.005}^{+0.005}$       \\
      & $d\Gamma_2$         &     $0.07_{-0.03}^{+0.03}$         \\
      & $E_{\rm br}$        &      $6.06_{-0.09}^{+0.10}$       \\  
      \hline
    $\chi^2/\nu$        &                                      &   $856.49/846=1.01$            \\
    \hline                                                    
    \hline
    \end{tabular*}
    \label{tab:bestfit_NS}
\end{table}

\section{Concluding remarks and future work} \label{sec:concl}
This paper presents an XSPEC table model that isolates the reflection spectrum produced by returning radiation from the accretion disk, excluding any contribution from the corona. When applied to the black hole XRB 4U~1630--47 in the soft state, the model provides a good fit to the data from three simultaneous NuSTAR and NICER observations. The best fits indicate a high black-hole spin $(a_* \sim 0.99)$, a disk inclination of $i \sim 43^\circ -54^\circ$, a mass-accretion rate of $\dot{M}_{\rm BH} \sim (0.13-0.25) \, \dot{M}_{\rm Edd}$, and a source distance in the $7.3-11.3 \, \rm{kpc}$ range, which is broadly consistent with previous studies. We also performed a simultaneous fit of the three observations, obtaining results consistent with a rapidly rotating black hole and favoring an inclination of approximately $47.5^\circ$ and a distance of about 9.6 kpc.

We compared our model with the widely used \texttt{relxillNS}. Our analysis shows that \texttt{relxillNS} requires a weak corona to adequately describe the data, in contrast to \texttt{zijiRetRad}. The latter produces a harder reflection spectrum at high energies and can  produce the high-energy tail without needing an additional Comptonized component. This behavior arises from the iterative reflection implemented in the \texttt{ziji} code, as well as from the high spin value inferred by our model (see Fig.~\ref{fig:it_model}).

Our model is, of course, a simplification, as we neglected the corona entirely, even though it is thought to be weak or even absent in soft states. Nevertheless, the results demonstrate that returning radiation alone may account for the observed spectra and should therefore be considered an important component during the high--soft state. We note, however, that the corona-free interpretation works best at the lowest flux, as the residuals at high energies in Obs. 3 may indicate the presence of a weak coronal component. A natural next step is to include a weak corona self-consistently in order to assess the relative contributions of coronal emission and self-irradiation. Extending the parameter grid of our model would also be valuable, particularly by including lower spin and mass values. Yet, as previously discussed, slower rotation would reduce the impact of returning radiation. Finally, to further generalize these findings, it will also be necessary to apply this approach to additional sources in the soft state.

\begin{acknowledgements}
K.K. would like to thank Arthur Suvorov and Marta Piscitelli for carefully reading the manuscript and providing valuable feedback. K.K. also acknowledges José Olvera Meneses for helpful discussions on accretion disk physics. The authors acknowledge support by the High Performance and Cloud Computing Group at the Zentrum für Datenverarbeitung of the University of Tübingen, the state of Baden-Württemberg through bwHPC and the German Research Foundation (DFG) for funding under "Project number 455787709" (bwForCluster BinAC 2).
\end{acknowledgements}

\bibliographystyle{aa} 
\bibliography{ref_AA} 

\begin{appendix}

\section{MCMC analysis for \texttt{zijiRetRad}}

To characterize the posterior distributions of the best-fit parameters listed in Table~\ref{tab:bestfit}, we employed the Markov chain Monte Carlo (MCMC) implementation in XSPEC, which uses the affine-invariant ensemble sampler of \cite{Goodman-2010}. To enable faster convergence, the calibration parameters for the \texttt{mbpo} and \texttt{gauss} models were fixed at their best-fit values. We verified this choice by running a long chain for Obs.~1 only, with all parameters free, and obtained consistent constraints on the parameters. We initialized 100 walkers and evolved the chains for 200,000 steps for each walker, discarding the first 5,000 steps as burn-in. We have verified that the chain length is at least 50 times larger than the autocorrelation length for each parameter. Furthermore, we compared the two-dimensional distributions for each pair of parameters between the first and second halves of the chain and found no significant differences.

\begin{figure*}[ht]
    \centering
    \includegraphics[width=\linewidth]{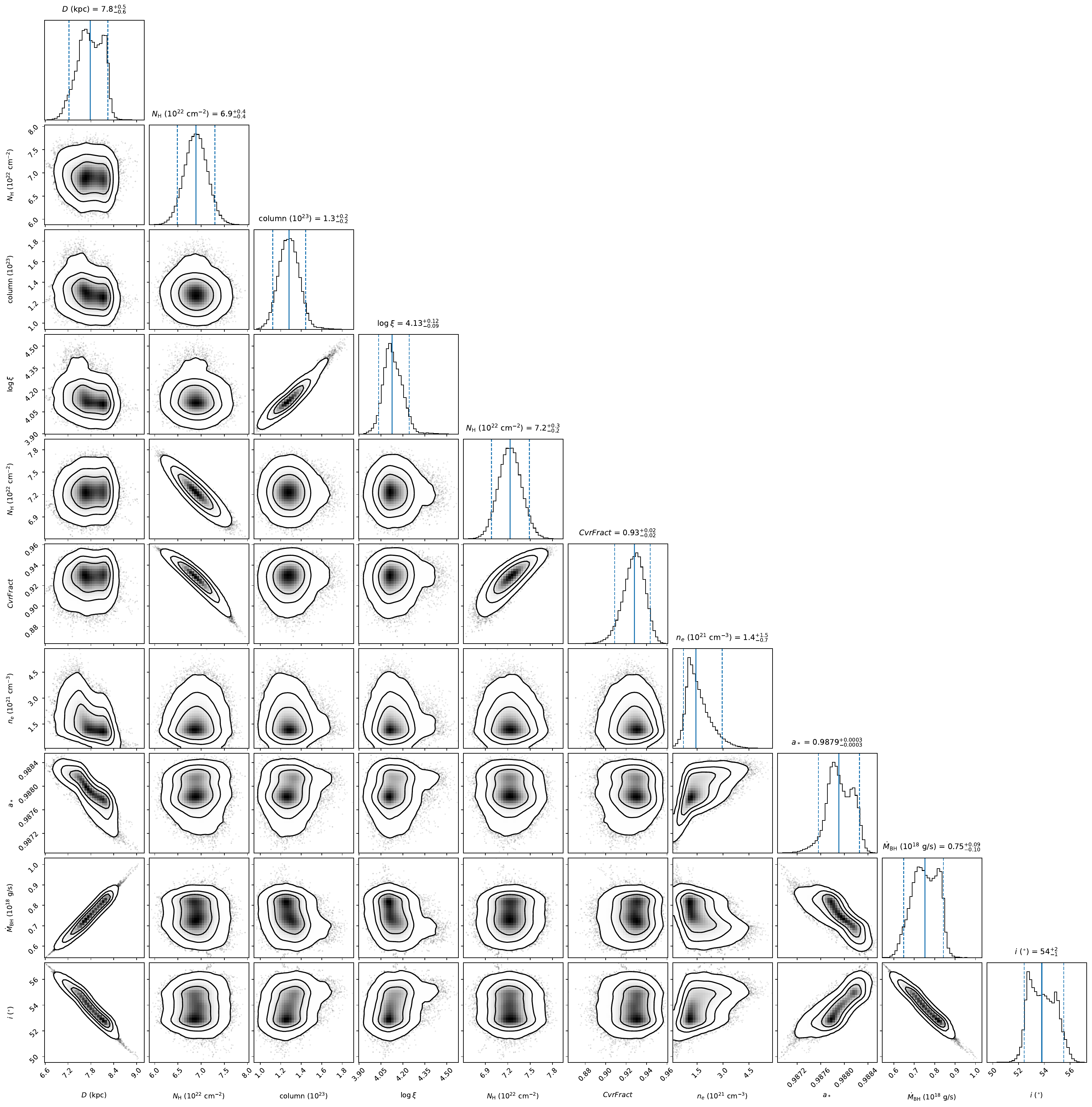}
    \caption{Corner plot for the best-fit model using \texttt{zijiRetRad} for Obs. 1. The error values of each of the parameters correspond to the $90\%$ confidence level. The contours correspond to the 68\%, 90\% and 99\% confidence levels.}
    \label{fig:MCMC_ni1}
\end{figure*}

\begin{figure*}[ht]
    \centering
    \includegraphics[width=\linewidth]{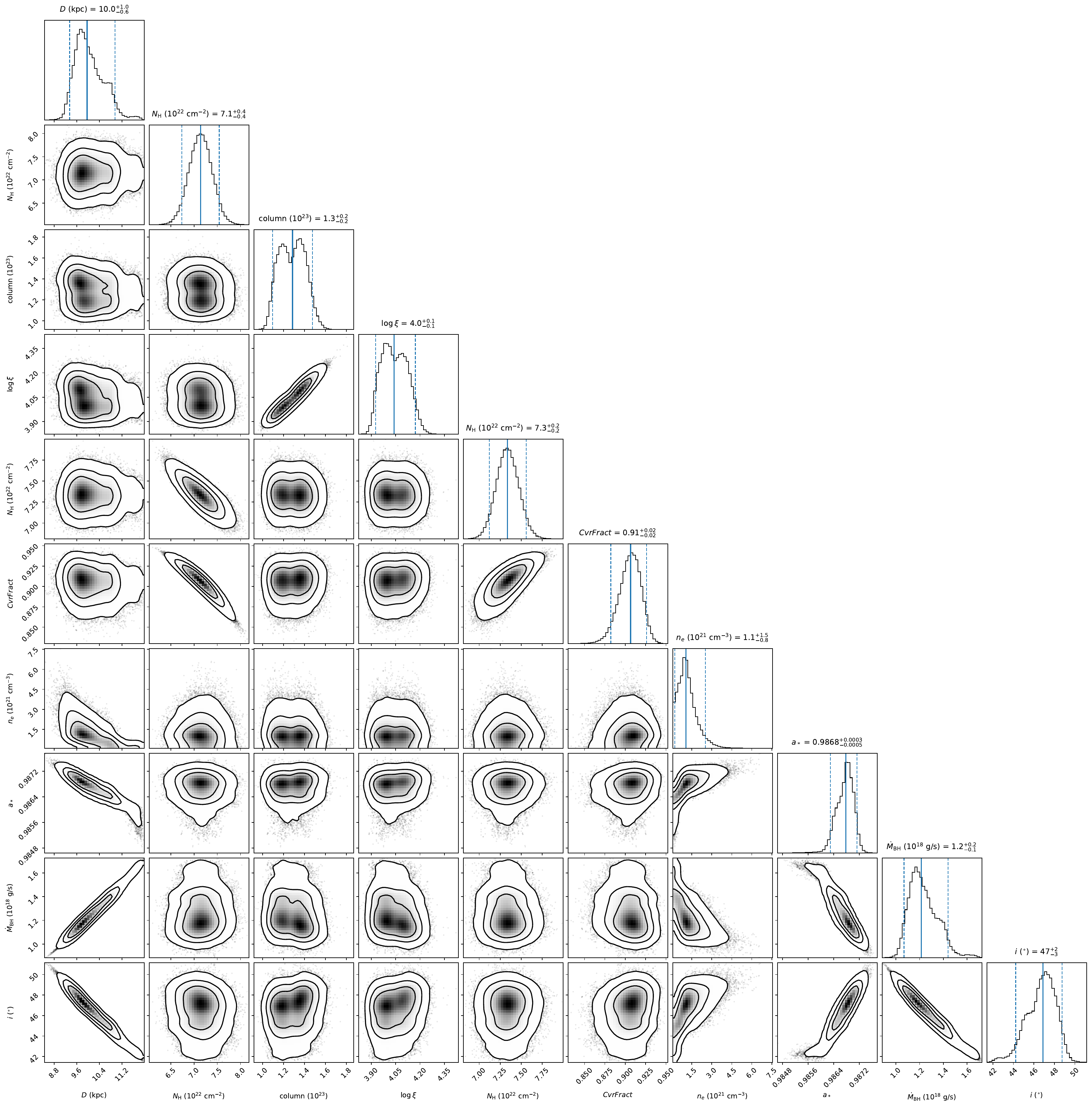}
    \caption{Same as Fig. \ref{fig:MCMC_ni1} but for Obs. 2. }
    \label{fig:MCMC_ni2}
\end{figure*}

\begin{figure*}[ht]
    \centering
    \includegraphics[width=\linewidth]{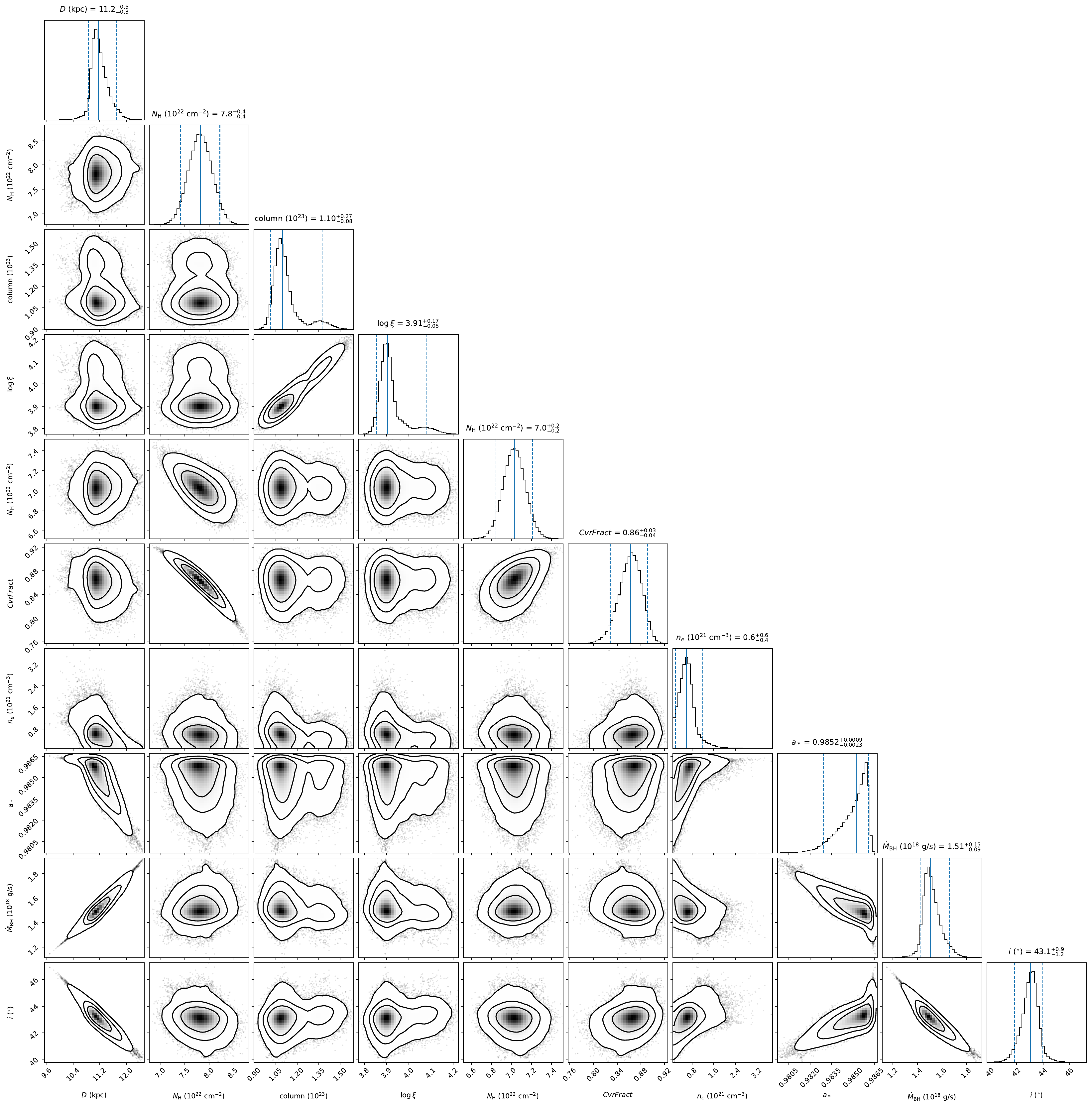}
    \caption{Same as Fig. \ref{fig:MCMC_ni1} but for Obs. 3. }
    \label{fig:MCMC_ni3}
\end{figure*}

The corner plots for the best fit model parameters are shown in Fig.~\ref{fig:MCMC_ni1}, ~\ref{fig:MCMC_ni2} and  ~\ref{fig:MCMC_ni3}. There is strong correlation between the two parameters of the \texttt{xstar} model; column density and ionization parameter. Physically, a higher 
ionization corresponds to  a larger fraction of fully stripped iron and therefore  a higher column density is required to produce the same absorption line. In the \texttt{zijiRetRad} model, the mass accretion rate and the source distance are also strongly correlated, since both primarily act as scaling factors for the observed flux, see Fig.~\ref{fig:ziji_model_plot}b.

\section{Distance estimation} \label{sec:dist}
The source distance can be inferred from the model's normalization through the flux-luminosity relation:
\begin{equation}
F_{\rm table}=\frac{L}{4 \pi D^2_{\rm ref}}, \quad    F_{\rm obs}=\frac{L}{4 \pi D^2_{\rm source}},
\end{equation}
and therefore:
\begin{equation}
    F_{\rm obs}=\left( \frac{D_{\rm ref}}{D_{\rm source}} \right)^2F_{\rm table}.
\end{equation}
Here, $F_{\rm table}$ is the flux at the reference distance $D_{\rm table}=10^6r_{\rm g}$ $(r_{\rm g}=GM/c^2)$ used in the table model and $F_{\rm obs}$ is the flux at the true source distance $D_{\rm source}$. Thus, the model's normalization plays the role of the geometric prefactor:
\begin{equation}
    \mathcal{N}=(D_{\rm ref}/ D_{\rm source})^2,
\end{equation}
so that $F_{\rm obs}=\mathcal{N}F_{\rm table}$. We derive the following relation, which directly connects the model's normalization to the physical distance:
\begin{equation}
    \mathcal{N}=0.229 \cdot 10^{-20}\, \frac{M^2 \, {\rm (M_\odot)}}{D^2 \, (\rm kpc)}.
\end{equation}
We note that the factor of $10^{-20}$ is absorbed into the calculated flux of the table model. Consequently, the normalization should be evaluated without this factor when using the model.

\end{appendix}
\end{document}